\documentclass[useAMS,usegraphicx,usenatbib]{mn2e}
\usepackage[a4paper,centering, totalwidth=520pt, totalheight=700pt]{geometry}
\usepackage{graphicx}
\usepackage{aas_macros}     
\usepackage{amsmath}
\usepackage{amssymb}
\usepackage{fixltx2e}[1999/12/01]
\usepackage{bm}

\newcommand{\erisbh}{Eris{\it BH} }

\newcommand{\Msun}{~\rm{M}_{\odot}}
\newcommand{\Lsun}{~\rm{L}_{\odot}}

\newcommand{\kms}{~\rm{km ~ s^{-1}}}
\newcommand{\pc}{~\rm{pc}}
\newcommand{\erg}{~\rm{erg}}
\newcommand{\ergs}{~\rm{erg/s}}

\newcommand{\Lbol}{~L_{\rm bol}}

\newcommand{\Mpc}{~\rm{Mpc}}
\newcommand{\kpc}{~\rm{kpc}}

\newcommand{\yr}{~\rm{yr}}
\newcommand{\Myr}{~\rm{Myr}}
\newcommand{\Gyr}{~\rm{Gyr}}
\newcommand{\atomcc}{~\rm{atom/cc}}

\newcommand{\Mvir}{M_{\rm vir}}
\newcommand{\Rvir}{R_{\rm vir}}
\newcommand{\MBH}{M_{\rm BH}}
\newcommand{\MBulge}{M_{\rm Bulge}}
\newcommand{\MDisk}{M_{\rm Disk}}

\newcommand{\epsj}{J_{c}/J_{z}}
\newcommand{\omegam}{\Omega_{\rm M}}
\newcommand{\omegab}{\Omega_{\rm b}}
\newcommand{\ho}{\rm{h}_{0}}

\title[Black Hole Evolution in a Late-type Galaxy] {Black Hole Starvation and Bulge
Evolution in a Milky Way-like Galaxy}

\author[]
{
Silvia Bonoli$^1$\thanks{E-mail:sbonoli@cefca.es},
Lucio Mayer$^2$, 
Stelios Kazantzidis$^3$, 
Piero Madau$^4$, 
Jillian Bellovary$^{5,6,7}$
\newauthor
and Fabio Governato$^8$\\
$^1$Centro de Estudios de F\'isica del Cosmos de Arag\'on, Plaza San Ju\'an 1,
Planta 2, 44001, Teruel, Spain\\
$^2$Center for Theoretical Astrophysics and Cosmology, Institute for
Computational Science, University of Zurich, \\
 Winterthurestrasse 190, 8057
Zurich, Switzerland \\
$^3$Section of Astrophysics, Astronomy and Mechanics, Department of Physics,
University of Athens, 15784 Zografos, Athens, Greece\\
$^4$Department of Astronomy \& Astrophysics, University of California, 1156 High Street, Santa Cruz, CA 95064\\
$^5$Department of Astrophysics, American Museum of Natural History, Central Park
West \& 79th St, New York, NY 10024\\
$^6$Department of Natural Sciences and Mathematics, Fisk University, 1000 17th
Avenue N., Nashville, TN 37208\\
$^7$Department of Physics and Astronomy, Vanderbilt University, PMB 401807,
Nashville, TN 37206\\
$^8$ Astronomy Department, University of Washington, Box 351580, Seattle,
WA 98195-1580
\vspace{-0.5cm} 
}

\begin{document}



\maketitle

\label{firstpage}

\begin{abstract}

We present a new zoom-in hydrodynamical simulation, ``\erisbh'' which follows the
cosmological evolution and feedback effects of a supermassive black hole at the
center of a Milky Way-type galaxy. 
\erisbh shares the same initial conditions, resolution, recipes of gas cooling, star
formation and feedback, as the close Milky Way-analog ``Eris'', but it
also includes prescriptions for the formation, growth and feedback of supermassive black
holes. The aim of this simulation is to study the evolution of supermassive
black holes and AGN feedback effects in a late-type galaxy.   We find that the galaxy's central
 black hole grows mainly through mergers with other black holes coming
from infalling  satellite galaxies. The growth by
gas accretion is minimal because very little gas reaches the sub-kiloparsec scales,
possibly  a reflection of the fact that spiral galaxies have a long period
of quiescent evolution in which their potential is only weakly perturbed by
external triggers. The final black hole is, at $z=0$, about 2.6 million solar
masses and it sits closely to the position of SgrA* on the 
 $\MBH-\MBulge$ and $\MBH-\sigma$ planes, in a location consistent with what observed for pseudobulges.  
Given the limited growth due to gas accretion, we argue
that the  mass of the central  black hole should be above $10^5 \Msun$ already
at $z \sim 8$. 
The effect of AGN feedback on the host galaxy is 
 limited to the very central few hundreds of parsecs. 
Despite being weak,  AGN feedback seems
to be responsible for the limited growth of the central bulge with respect
to the original Eris, which results in a significantly flatter rotation curve
in the inner few kiloparsecs. Moreover, the disk of \erisbh is more prone to
instabilities, as its bulge is smaller and its disk larger then Eris. As a result, the disk of \erisbh undergoes a stronger dynamical evolution
relative to Eris and around $z=0.3$ a weak bar grows into a strong bar of a few disk
scale lengths in size. The bar triggers a burst of star
formation in the inner few hundred parsecs, provides a modest amount of new
fuel to the central black hole, and causes the bulge of \erisbh to have, by
$z=0$, a box/peanut morphology.


\end{abstract}


\section{Introduction}


Present in nearly all spheroids \citep[e.g.,][]{kormendy04}, and being the
engines powering active galactic nuclei (AGN), supermassive black
holes (simply black holes, from here onwards) seem to be an integral component
of massive spheroidal galaxies.  
The tight relations between black hole mass and several properties of the host spheroid 
\citep[e.g.,][]{magorrian98,merritt01,tremaine02,haring04}  give
further hints that the life of black holes and their hosts are closely linked. 
Starting from the first analytical works on black hole ``self-regulation''
\citep[][]{silk98, king03}, a
considerable amount of effort has been spent in the last two decades in
understanding the role of black hole feedback in shaping the properties of the
host galaxy and controlling its own growth. Significant advance has been made  by
including the physics of black hole
accretion and feedback (even if limited to sub-grid models) into simulations of
galaxy evolution. For example, \citet{diMatteo05} and
\citet{springel05c} studied the effects of black hole (thermal) feedback during the merger
of blue spiral galaxies, and found that AGN feedback has a primary role in the color
transformation of the host galaxy, by
 quenching star formation and causing the elliptical merger remnant to redden in
a short timescale. The combination of these results and the evidence of scaling
relations between black hole mass and bulge properties, hint at the
idea that is during violent, major merger events when black holes acquire most of their
mass while elliptical galaxies form \citep[e.g.,][]{hopkins06}. On larger
scales, mechanical AGN
feedback in the form of jets and bubbles seems to halt cooling flows
in galaxy clusters \citep[e.g.,][]{churazov02,sijacki06}.
While black hole growth and the effects of AGN feedback have been
extensively explored in the case of massive  early-type galaxies in clusters
\citep[e.g.,][]{martizzi12,dubois13,martizzi14}, the role black
holes have played in the evolution of blue late-type galaxies evolving in
 smaller haloes is still largely
unexplored. 
The small interest the community has devoted to the interplay between black
holes and spiral galaxies is primarily due to the fact that late-type galaxies
have small bulges and  small black holes. Also, their active star formation
 suggests that no
quenching  is present in those galaxies, and thus AGN feedback is not significant. 

In the effort of simulating the evolution of disk galaxies, a lot of attention has
been instead
given to stellar feedback. Obtaining realistic late-type
galaxies has been a historical challenge for computational astrophysics since the first
simulations run in the early nineties. \citet{navarro91} found that the baryonic component of galaxies evolving
in hierarchically growing haloes
 was not able to retain enough angular momentum to feature a cold disk and flat
rotation curves as in realistic spirals. Those authors
 suggested that a proper treatment of supernova feedback (or some other
form of heating) at early times
 would be necessary to prevent gas to lose angular momentum 
and  catastrophically cool during galaxy interactions. Since these first
simulations, our understanding of the formation of spiral galaxy has evolved
significantly and, in the last few years,  a large number of works from
different groups has
shown that proper resolution and an
accurate treatment of star formation and stellar feedback are the key ingredients for creating realistic late-type galaxies
\citep[e.g.,][]{guedes11,brook11,aumer13,okamoto13,roskar14,marinacci14,agertz15,murante15}.
For further
details on this topic, we refer the
reader to the early review of \citet{mayer08}, the large code-comparison work of
\citet{scannapieco12} and the summary of all most recent developments
given in section 2 of \citet{murante15}. 
We discuss here only the work of \citet{guedes11}, as that is the
starting point of the new simulation presented in this paper. \citet{guedes11}
performed a zoom-in cosmological hydro simulation of a Milky Way-size halo with
a rather quiet merger history. They found that the combination of
high spatial resolution (gravitational softening of $120 \pc$) and  the use of a
high-density threshold for star formation \citep[as first used by][]{governato10} and of blast-wave feedback
\citep[as in][]{stinson06}, contribute to the development of a clumpy and inhomogeneous
 interstellar medium, where  overlapping SN explosions are able to inject enough
energy to remove low-angular momentum material. The simulation of
\citet{guedes11}, dubbed ``Eris'', is one of the first successful efforts to
produce a realistic late-type spiral in a cosmological simulation.

With the new simulation that we present in this paper, ``\erisbh'', we want to
examine directly the role of AGN feedback in the evolution of late-type galaxies 
and explore what could be the origin and cosmological evolution of the black holes
that these galaxies host.
We do so by adding sub-grid physics for the growth and feedback of supermassive
black holes to the physics already simulated in Eris.
The paper is organized as follows. Section \ref{sec:simulation} offers a brief
summary of the Eris simulation and describes the  technical aspects of 
\erisbh. In Section \ref{sec:BHs} we  show the evolution of
the black holes in the simulation, focusing in particular on the  central black
hole of the main galaxy of the simulation. In Section \ref{sec:Galaxy} we show how, and to which extent, AGN
feedback influences the host galaxy, by making a direct comparison between the
 physical properties of \erisbh and those of Eris. Finally, in Section
\ref{sec:conclusions} we summarize of our results.

\section{The simulation} \label{sec:simulation}

We describe here the technical aspects of the simulation. The initial conditions
and the physical processes included in the simulation are the same as the ones
of Eris, which are described in Section \ref{sec:Eris}. In addition, this
new simulation also includes assumptions for the formation of massive black
holes and follows their growth as described in Section \ref{sec:ErisBH}.

\begin{figure*}
\begin{center}
        \includegraphics[width=0.36\textwidth]{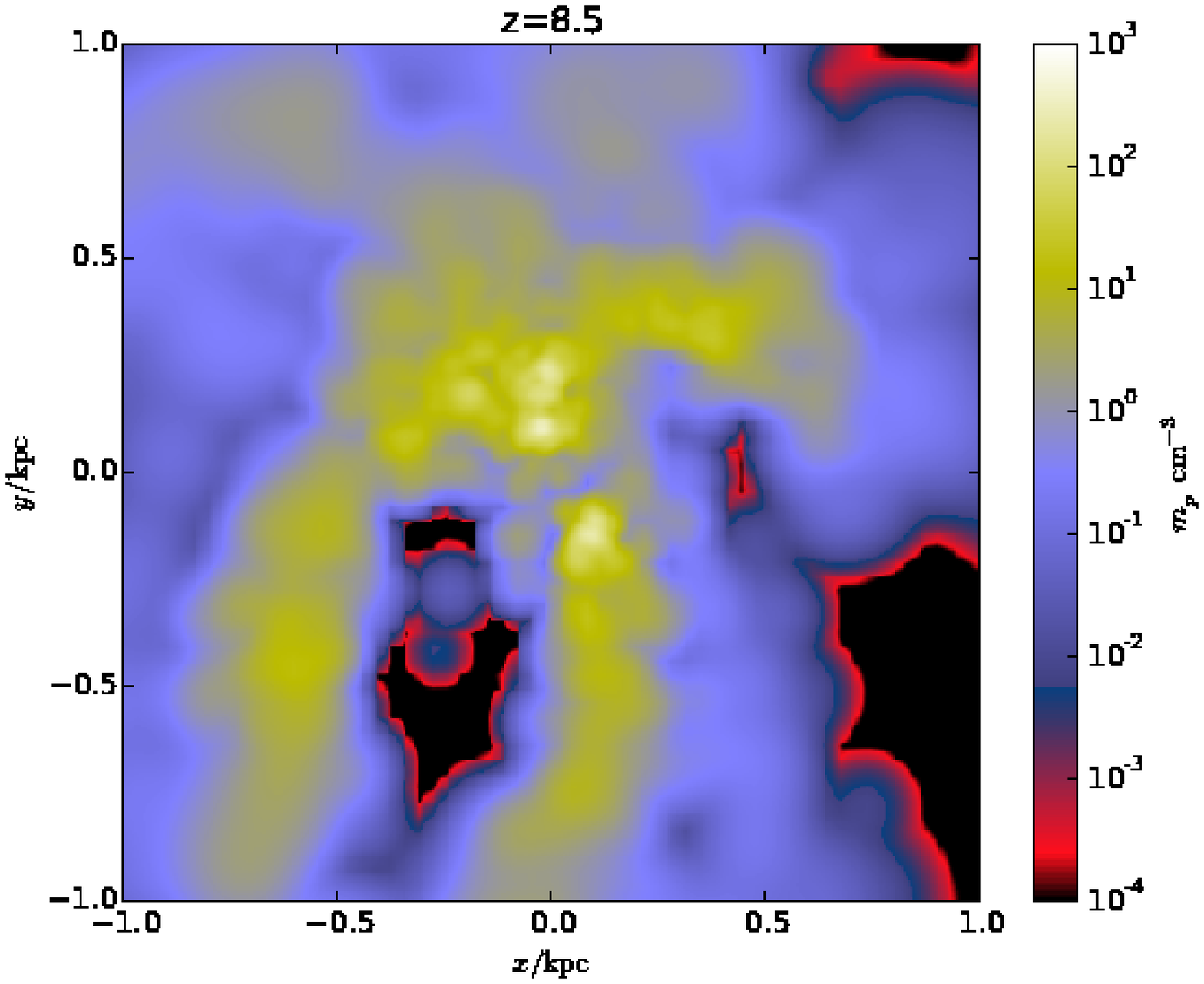}
        \includegraphics[width=0.36\textwidth]{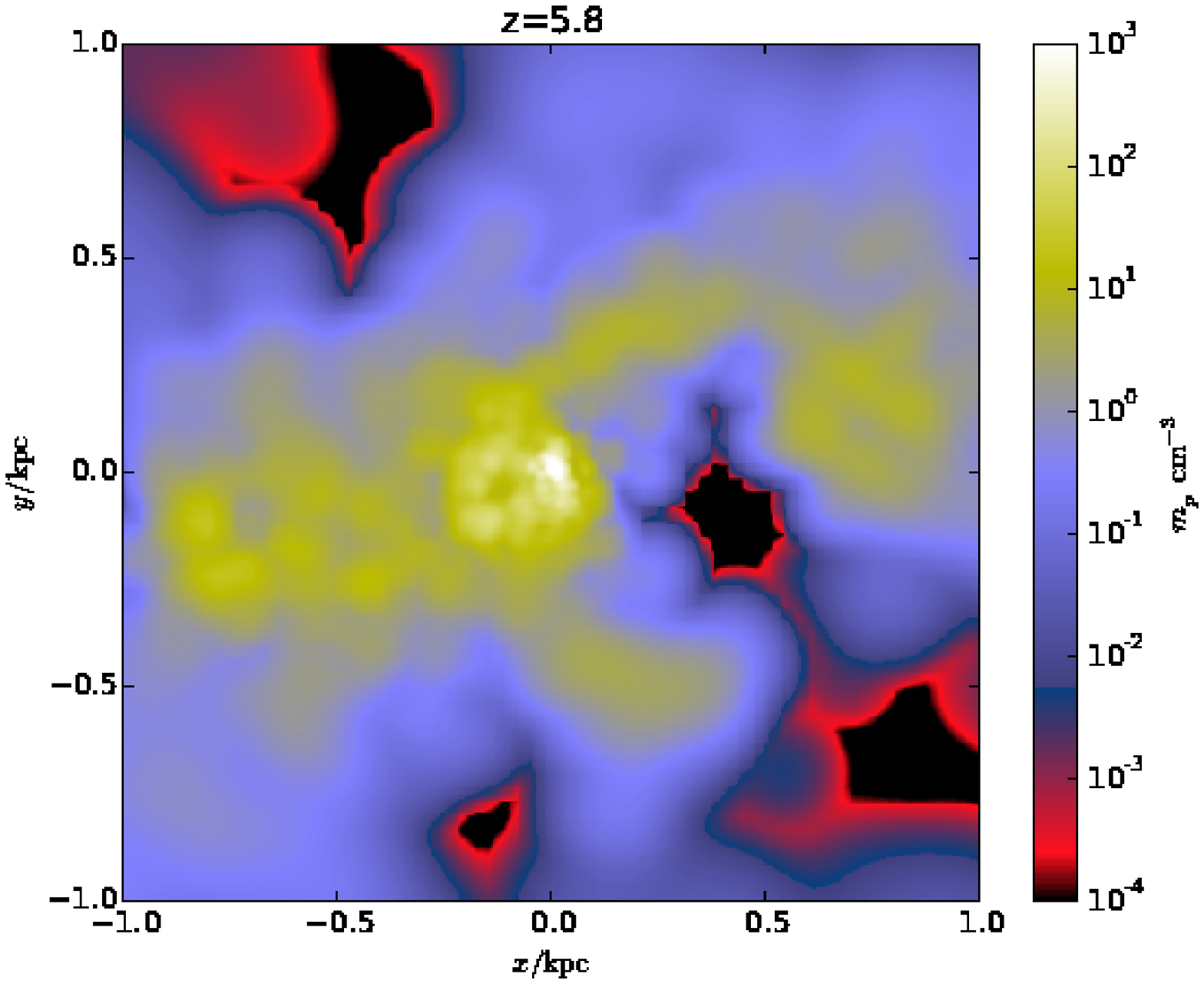}
        \includegraphics[width=0.36\textwidth]{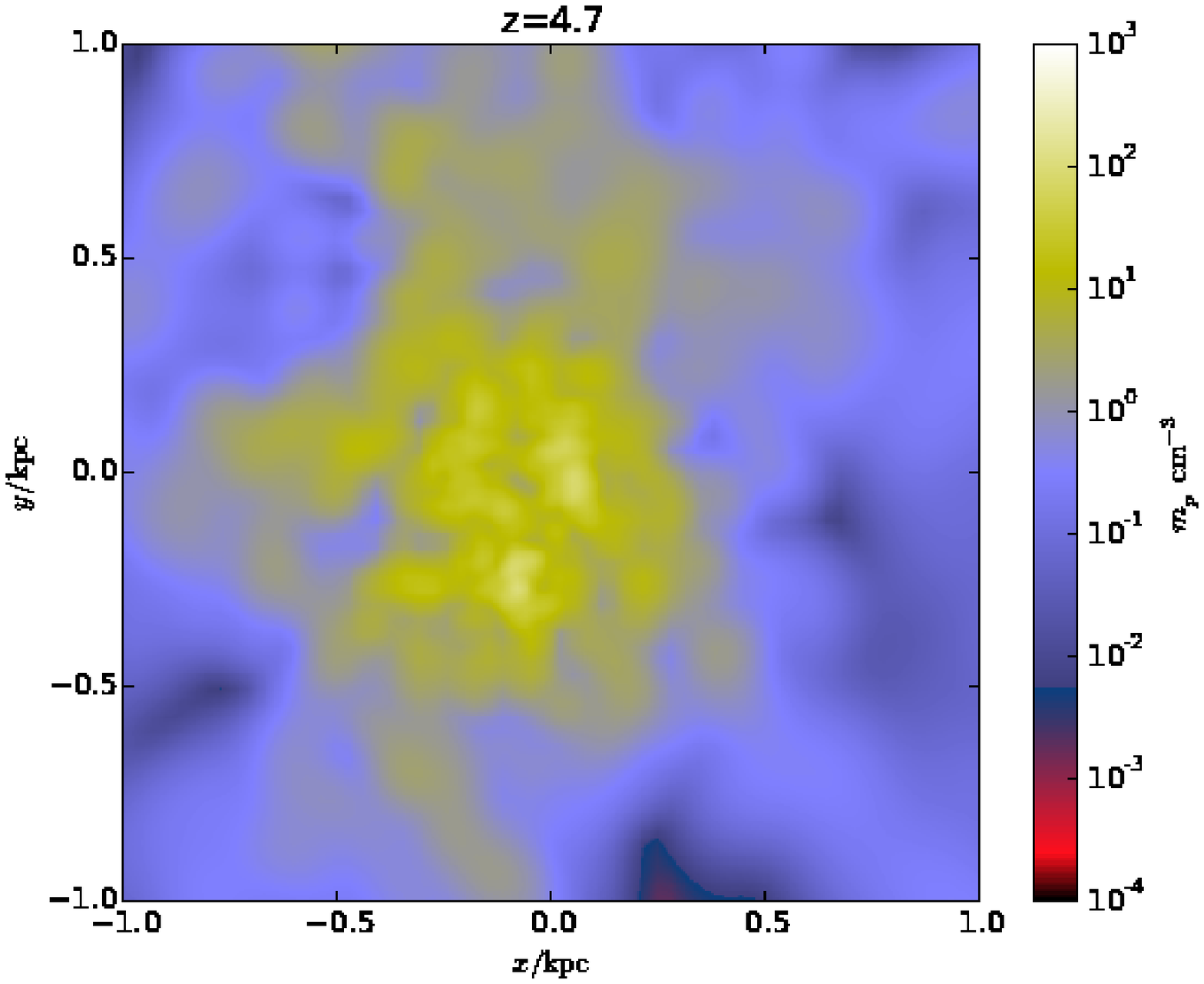}
        \includegraphics[width=0.36\textwidth]{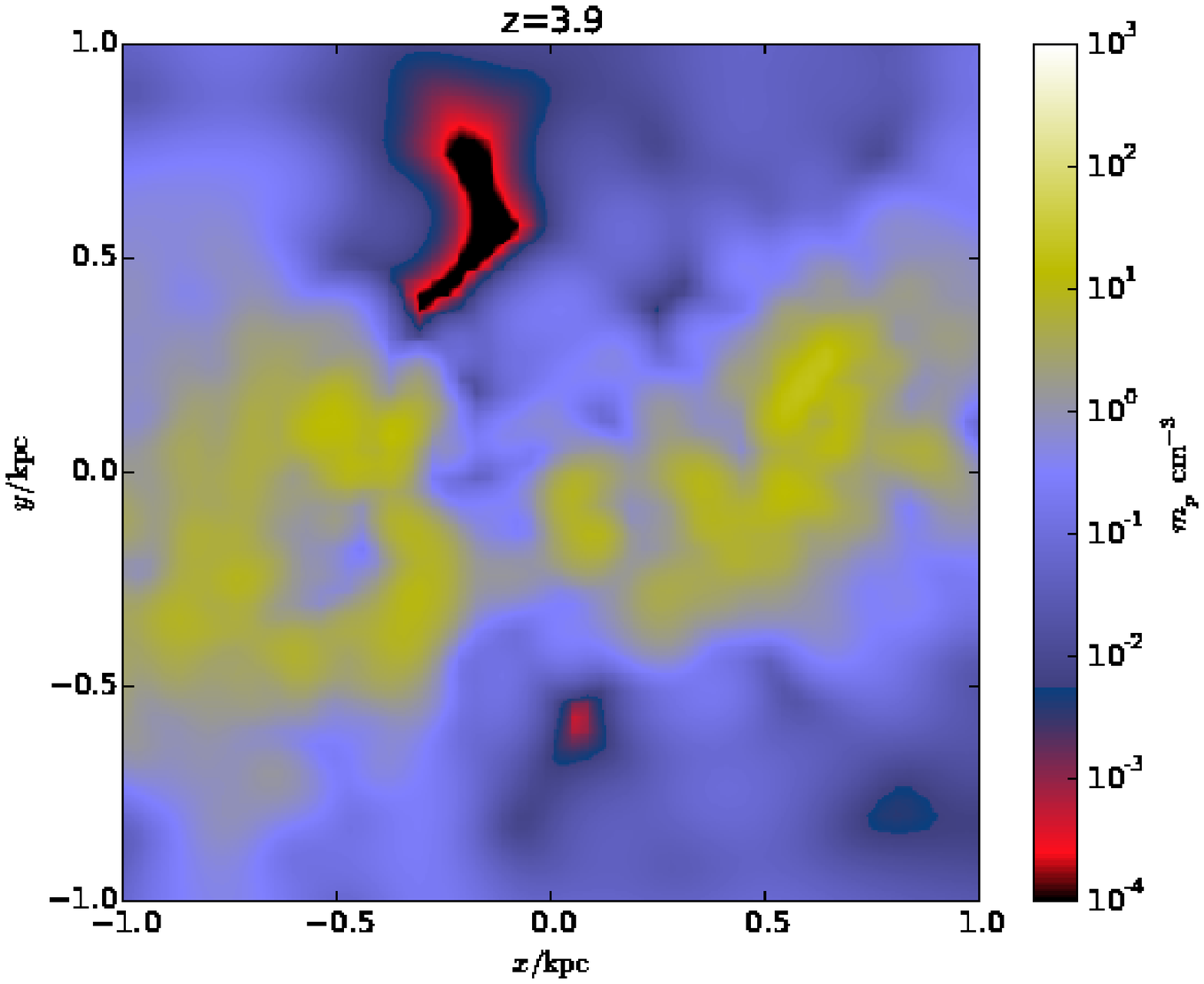}
        \caption{Gas density maps in the location where new black hole seeds are
about to be inserted in the simulation.}
        \label{fig:map_seed}
\end{center}
\end{figure*}

\subsection{Eris} \label{sec:Eris}

Eris is a cosmological zoom-in $N$-body/smooth particle hydrodynamic (SPH)
simulation that follows the evolution of a Milky Way-size halo from $z=90$ down to
$z=0$, in a {\it Wilkinson Microwave Anisotropy Probe} three-year cosmology
\citep{spergel07},
with a flat universe with $\omegam=0.24$, $\omegab=0.042$, $\ho=73
\kms\Mpc^{-1}$, $n=1$ and $\sigma_{8}=0.76$. It was run with the parallel, spatially
and temporally adaptive, treeSPH-code {\it GASOLINE} \citep{wadsley04} for 1.5
million cpu hours. The halo was selected from the $z=0$ output of a low-resolution and dark matter-only
simulation of a $(90  \Mpc)^3$ volume; it was chosen to have approximately
the estimated size of the Milky Way halo\footnote{The virial mass if Eris at $z=0$ is
$\Mvir = 7.8 \times 10^{11}\Msun$, and the virial radius $\Rvir = 234 \kpc$, as
given by the halo finder {\scshape AMIGA} \citep{gill04, knollmann09}.} and to
have had a rather quiet merger history, with no major mergers (above the ratio of
$1:10$) after $z=3$. The dark matter and the (initial) gas particle masses of the
high-resolution region of Eris are  $m_{\rm DM} = 9.8 \times 10^{4} \Msun $ and
$m_{\rm gas} = 2 \times 10^4 \Msun$. The gravitational softening length was
fixed to $120$ physical $\pc$ from $z=9$ to the present, and evolved as
$1/(1+z)$ from $z=9$ to the starting redshift ($z=90$). 
In brief \citep[for further technical details, see][]{guedes11}, the simulation
includes Compton cooling, atomic cooling,
metallicity-dependent radiative cooling at low temperatures and uniform
UV background. Star formation and stellar feedback are regulated by the star
formation threshold $n_{\rm SF}$ (set to a value of $5 \atomcc$), the star formation
efficiency $\epsilon_{\rm SF}$ (set to $\epsilon_{\rm SF}=0.1$) and the
fraction of of supernova energy that couples to the interstellar medium
$\epsilon_{\rm SN}$ (set to $0.8$). When the conditions of the gas become
favourable to star formation (according to the Schmidt law), new star particles
are created following the \citet{kroupa01} initial mass function,  with a
starting  mass $m_{*}=6\times 10^3 \Msun$ (the gas particle from which the
star is created has its mass reduced by the same amount). Supernova explosions imply a deposition
of metals and energy ($\epsilon_{\rm SN} \times 10^{51} \erg$) to the nearest
neighbour gas particles. The affected gas has its cooling shut off until the
end of the supernova blastwave, as in \citet{stinson06}.
Winds and stellar mass loss are also modelled as in \citet{stinson06}.

As discussed in \citep{guedes11},  a high value for the density
threshold for star formation parameter was possible thanks to the high mass and spatial
resolutions of the simulation, which allow to resolve the clouds where star
formation occurs. With this choice, star formation takes place in confined
regions, giving rise to a clumpy, inhomogeneous interstellar medium, where
overlapping supernova explosions inject energy in a localized manner. This
localized energy injection is able to create galactic outflows which expel
low-angular momentum material. 

The formation of a clumpy and inhomogeneous medium  seems  to be the key ingredient to form a realistic
late-type spiral. As described in \citep{guedes11}, in fact, Eris is consistent
with a large number of observational aspects of the Milky Way, from  its
structural properties to the mass content of
its different components. Just to give few examples, it has
a flat rotation curve, a  low  photometric bulge-to-disc (B/D) ratio, it falls on
the Tully-Fisher and stellar-mass/halo-mass relation and it has a baryonic
mass fraction within the virial radius which is $30\%$ lower than the cosmic
value. 

\begin{table*}
 \begin{center}
  \begin{tabular}{c||cc|cc|cc|cc|}
	\hline
	  & z - seed & $N_{\rm gas} > 100$ atom/cc & $M_{\rm vir}$ &  $M_{\rm gas}$ &
$M_{\rm star}$ & $M_{\rm BH seed}$\\
        \hline
        \hline
        $1$ & $8.5$ & $1354$ & $7.2 \times10^9 \Msun$ & $7.2\times10^8 \Msun$  & $1.7 \times10^8 \Msun$ & $8.7 \times10^5 \Msun$ \\
        \hline
        $2$ & $ 5.8$ & $579$ &  $4.8 \times10^9 \Msun$ & $5.9\times10^8 \Msun$  & $1.7 \times10^8 \Msun$ & $1.3 \times10^5 \Msun$ \\
        \hline
        $3$ & $ 4.7$ & $318$ &  $7.0 \times10^9 \Msun$ & $1.4\times10^9 \Msun$  & $2.0 \times10^8 \Msun$ & $7.6 \times10^5 \Msun$ \\
        \hline
        $4$ & $ 3.9$ & $13$ &  $6.4 \times10^9 \Msun$ & $7.0\times10^8 \Msun$ & $1.2 \times10^8 \Msun$ & $0.8 \times10^5 \Msun$ \\
        \hline
  \end{tabular}
   \caption{Summary of the physical properties of the systems
 that qualify as proper sites for black hole seeding, following the
conditions described in Section \ref{sec:seeding}. The first column gives the
redshift at which the black hole is inserted in the system. The second column
 gives the number of gas particles that have a density higher than $100$
atom/cc. The remaining columns give the virial, gas and stellar mass of the
systems as well as the mass of the black hole seed.}
  \label{table:info_seed}
 \end{center}
\end{table*}

\subsection{\erisbh: Eris {\it plus} black hole formation, growth and feedback}
\label{sec:ErisBH}

\erisbh is a replica of Eris, but with additional prescriptions for the seeding
and  growth of supermassive black holes,  and  thermal
feedback during gas accretion.

\subsubsection{Seeding procedure}  \label{sec:seeding}

The origin of black hole seeds is still largely a mystery and subject of intense
theoretical investigation \citep[see, e.g., the review of][]{volonteri12}. Thus, 
the criteria for  inserting black hole seeds in a  cosmological  simulation
 is somewhat arbitrary. \citet{diMatteo08} and \citet{booth09}, for example,  have chosen
to insert a seed
black hole in every dark matter halo that rises above a given threshold in
 mass. 
\citet{bellovary10}, instead, chose to connect the formation of  black
holes to star formation: gas particles with zero metallicity and with density
above  the set
threshold for star formation, have a certain probability to be transformed into
black holes rather than stars, where this probability parameter is tuned to
reproduce the black hole seed halo occupation probability at $z=3$ suggested
by \citet{volonteri08}  (which, however, is based on a very
specific formation model for black hole seeds).
Here, we impose both a requirement on resolution and high-density gas
environment for the introduction of new black holes. A seed is inserted in
all systems that satisfy the following conditions:
\begin{itemize}
\item{are bound, according to the {\scshape AMIGA} halo finder
\citep{gill04, knollmann09}, and resolved with at least  $10^5$ particles,}
\item{have  at minimum of $10$ gas particles with density
above $100~ \rm{atoms/cc}$. }
\end{itemize} 
The resolution criterion is motivated by past works on numerical convergence
of angular momentum transport in astrophysical disks simulated by SPH,
which have shown that at least $10^5$ particles are needed to keep a number of
numerical
effects under control, such as spurious hydrodynamical drags as well as
enhanced gravitational torques from a noisy halo potential \citep{kaufmann07}.

Galaxies that satisfy the conditions listed above are then seeded with a black
hole,
provided that they do not host one already.
This is done by selecting the star  particle that
is closest to the high-density gas particle with the deepest potential and 
 converting it into a sink particle. The
initial black hole mass
is set to be proportional to the number of high-density gas
particles\footnote{specifically, we calculate the number $N_{gas,dens}$ of gas
particles that have a
density larger than $50 \%$ the density of the most dense gas particle in the
galaxy. The mass of the black hole seed is then given by $N_{gas,dens}$ times
the mass of the converted star particle.}, so that
the  higher  is the density of the gas, and the larger is  the high-density
region, the more massive is the newly-formed black hole. 

From the beginning of the simulation down to $z \sim 3$, only six structures
raise above the mass (or number of particles) limit that we have imposed. Of
those, four satisfy also the condition on the gas density. At lower redshift,
the gas densities become generally lower, and we stop looking for possible sites for new
black holes seeds.

The first
 system which satisfies the seeding conditions is, at $z \sim 8.5$, the
progenitor of the main galaxy of the simulation. The other
three seeds are inserted in satellite galaxies at lower redshifts. Figure
\ref{fig:map_seed} shows maps of the gas density in the regions where new seeds are
about to be added to the simulation.  Table~\ref{table:info_seed} gives further information on the
properties of the systems where black holes are inserted. Given the imposed limit
on the number of particles that a protogalaxy needs to have to host a black
hole, the masses in gas, star and dark matter do not differ substantially in the
four galaxies at the time each black hole is inserted. Seed masses (last column
in the table) also do not
differ significantly, with about one order of magnitude separating the most to 
the
least massive seed. As a consequence, all black hole seeds are located close to each
other on the $\MBH -$ stellar mass plane, slightly above the extrapolation to
lower masses of the relation established by the most massive black holes (see
left panel of Figure \ref{fig:MBH_star}).

\subsubsection{Black hole accretion and feedback model} \label{sec:acc_feedback}

After being inserted into the simulation by converting the appropriate star
particle into a sink-type particle, black holes start
accreting gas isotropically from the surrounding medium, following the widely used
Bondi-Hoyle-Lyttleton formalism \citep{hoyle39,bondi44, bondi52}:
\begin{equation}\label{eqn:bondi}
\dot{M}_{\rm Bondi}=\frac{4\pi G^2M^2_{\rm BH}\rho}{(c_{s}^2+v^2)^{3/2}},
\end{equation}
where $\rho$ and $c_s$ are, respectively, the density and sound speed of the
gas, $M_{BH}$ is the mass of the black hole, and $v$ is the velocity of the
black hole relative to the gas.  Growth proceeds according to
Equation \ref{eqn:bondi}, but the maximum allowed accretion rate is set to the  
Eddington accretion rate $\dot{M}_{\rm Edd}=(4 \pi G M_{BH}
m_{p})/(\eta \sigma_{T} c)$ (where $m_p$ is the proton mass, $c$ is the speed of
light, $\sigma_T$
the Thomson cross-section and $\eta$ is the accretion efficiency, 
assumed to be $0.1$), so that:
\begin{equation}\label{eqn:Mdot_BH}
\dot{M}_{\rm BH}=
\left\{
\begin{array}{lr}
\dot{M}_{\rm Bondi} &\mbox{if } \dot{M}_{\rm Bondi}<\dot{M}_{\rm Edd} \\
\dot{M}_{\rm Edd} &\mbox{if } \dot{M}_{\rm Bondi}>\dot{M}_{\rm Edd} \\
\end{array}
\right. 
\end{equation}
 Feedback from the
accreting black hole is modeled by assuming that a fraction $\epsilon_{f}=0.05$
of the radiated luminosity\footnote{The bolometric radiated luminosity is given
by $L_{\rm Bol}=\epsilon_{\rm rad} \dot{M}_{\rm BH} c^2$, where $\epsilon_{\rm
rad}$ is the radiative efficiency. For
simplicity, we assume here $\epsilon_{rad}=\eta$, which is a good approximation
 at high accretion rates when black holes are likely growing from
geometrically-thin and optically-thick accretion disks \citep{shakura73}, but it
might be an overestimate of the radiative power when accretion rates are highly
sub-Eddington \citep{churazov05, merloni08}.}
 is converted into thermal energy that heats the gas
surrounding the black hole.  Initially used by
\citet{diMatteo05} and
\citet{springel05a} in their study of the effects of black hole growth and
feedback in isolated galaxy mergers, the Bondi-prescription for estimating the
growth of massive black hole has been used widely in the community, also in
large cosmological volumes \citep[e.g.][]{diMatteo08,booth09} (we discuss the
limitation of the Bondi-Hoyle-Lyttleton formalism below in section
\ref{sec:discussion_bondi}).
\citet{bellovary10} have included this prescription into the GASOLINE code to
study “wandering” black holes, that is, black holes that are the
remnants of stripped satellite cores. 
As in \citet{bellovary10}, black holes are allowed to merge if they are within one another’s
softening length and if they fulfill the criterion $\frac{1}{2} \Delta v^2 <
\Delta a \dot \Delta r$,
where $\Delta v$ and $\Delta a$  are the differences in velocity and acceleration
of the two black holes, and $\Delta r$ is the distance between them.

\begin{figure}
\begin{center}
        \includegraphics[width=0.49\textwidth]{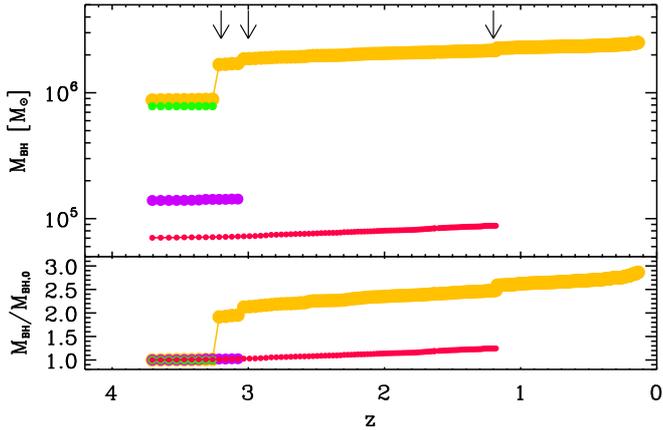}
	\caption{Upper panel: mass evolution of the four black holes in the simulation.
Symbols stop when the corresponding black hole merges with the black hole of the
 central galaxy (yellow track). Lower panel: again mass evolution of the four black
holes, but now normalized by the respective initial mass. The vertical arrows
indicate the redshift at which black hole mergers take place. }
	\label{fig:BHevol}
\end{center}
\end{figure}

\section{Black hole properties} 
\label{sec:BHs}

\subsection{Black hole growth} \label{sec:accretion}

The upper panel of Figure \ref{fig:BHevol} shows the mass evolution of the four black holes in the
simulation. Curves stop when the corresponding black hole merges with the black
hole of the central galaxy (yellow symbols). The mergers (marked by the arrows) are clearly visible as
``jumps'' in the yellow track. 
The lower panel of the figure shows again the growth of the four black
holes, this time normalized by the corresponding initial mass. Growth by gas
accretion is clearly limited in all black holes. The central black
  approximately doubles
its mass though mergers and, since the time it was seeded into the simulation at
$z\sim8.5$, it grows,
in total, only by a factor of $\sim 3$, reaching  a final mass of about $2.6$ million Solar masses.
 The gas growth we find in our simulation is much more modest compared to the
one obtained by 
\citet{marinacci14}  in their simulations of eight Milky Way-sized haloes:  all
 black holes hosted by their simulated central galaxies reach final
  masses around $10^8 \Msun$. We argue that the differences in the 
results are likely primarily due to the different resolutions adopted in our and
their simulations: \citet{marinacci14} have a gravitational softening of about
$700 \pc$, while we resolve down to $120 \pc$.  We argue that the lower
resolution of their simulations implies an overestimate of the gas supply
available for accretion. In our
simulation, for example, the amount
of gas inside $\sim 1 \kpc$ is around $10^8 \Msun$,  almost an order of
magnitude higher than in the inner $200-300 \pc$.

\begin{figure}
\begin{center}
        \includegraphics[width=0.49\textwidth]{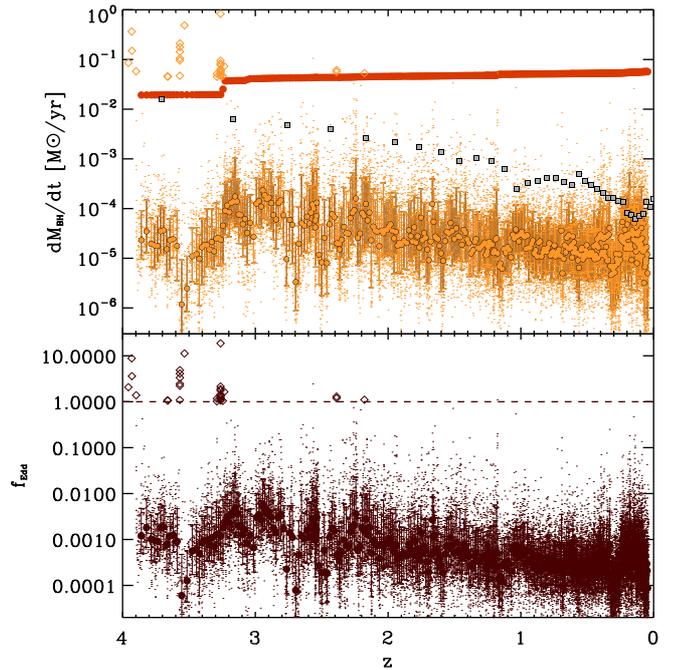}
	\caption{Upper panel: evolution of the instantaneous accretion rate of the black hole in the main
galaxy (orange points) with median values over multiple timesteps (orange bullets)
and $16$ and $84$ percentiles (error bars). 
The red curve indicates what would be the Eddington
rate for the black hole at each time. The orange diamonds highlight when the
 Bondi formula would have given a super-Eddington rate, given the gas properties around
the black hole. The grey squares show the evolution of the sfr (multiplied by
$10^{-3}$) in the inner $2
\kpc$ of the galaxy. Lower panel: evolution of the Eddington
ratio, with again median values over many timesteps and   $16$ and $84$
percentiles (filled bullets with error bars). The dashed line indicates the
Eddington limit and, again, the
diamonds show when the accretion rates would have been super-Eddington.}
   	\label{fig:BHmdot_sfr}
\end{center}
\end{figure}

Figure \ref{fig:BHmdot_sfr} shows the gas accretion rate evolution of the central
black hole. The orange points in the upper panel indicate the instantaneous
accretion rate, while the
orange bullets show the median over several timesteps (the error bars indicate the
$16$ and $84$ percentiles of the distribution). The median accretion rates 
 decrease only gently with redshift, and are typically quite low, $\sim
10^{-4} - 10^{-5} \Msun/\yr$, but there are periodic fluctuations and peaks at
higher accretion rates. While there is a global gentle decrease in the median
accretion rates with decreasing redshift, at $z \sim 0.3$ the median rate
increases again of about an order of magnitude: as we will discuss in the next
section, this is approximately the redshift at which the disk
of the galaxy is experiencing an instability event that leads to the formation
of a strong bar visible in the galaxy at $z=0$. During the process of bar
formation, the gas mass and the star formation increase
in the very central few hundred
parsecs  and the conditions around the black hole also
become favourable for an increase of the accretion rate. By $z=0$, the accretion
rate has lowered again to values around $10^{-5} \Msun/\yr$.

 In the same Figure, the red bullets indicate, for reference, which would be the instantaneous Eddington
accretion rate for the black hole.  We also explicitly show when the conditions of the gas around the
black hole would have given a Bondi accretion rate higher than the Eddington
limit (orange diamonds): this happens rarely, and only at high redshift.
However, as we
do not allow super-Eddington accretion, the accretion rate in those rare cases
is set to the Eddington limit.
 
The bottom panel explicitly shows the evolution of the Eddington ratio
($f_{Edd}= \dot{M}_{BH}/\dot{M}_{Edd}$), which is typically between $10^{-4}$ and
$10^{-2}$.  Again, the diamonds show the few rare times where the Bondi 
formula would give  accretion rates higher than the Eddington limit. By $z=0$
the Eddington ratio reaches the lowest median values of $\sim10^{-4}$. 

Accretion rates between $10^{-5}$ and $10^{-3} \Msun/\yr$ are, conservatively,  the accretion
rates expected from Bondi accretion of hot gas and stellar mass loss in the
center of nearby galaxies, as discussed by \citet{ho09}. At $z=0$ our simulated
black hole is accreting close to those values, at $\sim 10^{-5} \Msun/\yr$, which, assuming a radiative 
efficiency $\epsilon_{\rm rad}=0.1$, corresponds to a bolometric 
luminosity of $\sim 10^7 \Lsun$. This is close to the median value of 
 nuclear luminosity of local Seyfert galaxies found by \citet{ho09}. As
discussed in Section \ref{sec:acc_feedback} however, 
 the radiative efficiency is likely lower than the $10 \%$
value we have adopted here, and can be as low as $\epsilon_{\rm rad}\sim
10^{-3}-10^{-4}$ for the accretion rates we
have in our simulation at late time. At those low rates, in fact, energy from the accreting black
hole is likely released as kinetic, rather than radiative, power
\citep{churazov05, merloni08}. With significantly lower 
efficiencies of energy release, the luminosity of the nucleus of \erisbh could
be several order of magnitude lower, thus being closer to the luminosities that
\citet{ho09} find for a 
 large
fraction of local galaxies (including  the Milky Way\footnote{The estimated
luminosity of the Galactic
center is $\Lbol <10^{37} \ergs$
 \citep[see,e.g.][]{skinner87, pavlinsky94, baganoff03} and other references in
\citet{narayan98}.}). Had we assumed a lower value for the radiative efficiency for the
phases at the 
lowest Eddinton ratios, the effect of radiative AGN feedback on the galaxy would have been
even lower  than what we have obtained in our simulation.


In the top panel of Figure \ref{fig:BHmdot_sfr} we also show the evolution of
the star formation rate (multiplied by $10^{-3}$ for plotting purposes, and with
values
averaged over $10$ snapshots, which correspond to about $300 \Myr$) in the inner $2 \kpc$ of the
galaxy. As the bulge stars dominate approximately the inner $2 \kpc$ of the
galaxy at $z=0$ (see section \ref{sec:star}), and as most bulge stars are formed
in-situ \citep[see][]{guedes13}, this sfr is a good proxy for the
sfr of the stars within the bulge.  
The black hole accretion rate is between $6$ and $4$ orders of magnitude smaller
than the star formation rate. As expected for a  
late-type galaxy, the growth rate of the central black hole is thus significantly
smaller than the growth rate of the bulge. In section \ref{sec:scalings} we 
explicitly show how this translate into black hole-galaxy scaling relations.

Given the little gas supply, the value of the initial seed mass is
rather important. Generalizing our results, we expect the progenitors of Milky Way-size
galaxies to be hosting an intermediate-mass black
hole already at $z \sim 8$. For seeds from PopIII remnants
\citep[e.g.,][]{bond84, madau01, haiman01, tanaka09}, given the
little time available before $z \sim 8$, uninterrupted growth close to the
Eddington rate would be required to reach intermediate masses \citep[unless
super-Eddington accretion is possible, see, for
example,][]{wyithe12,madau14,volonteri15}. An alternative
are seeds from
direct-collapse, which could be already between $10^4$ and $10^6 \Msun$ at the
time of formation. Given the properties of Milky Way progenitors, direct collapse black holes born
in metal-free protogalaxies \citep[e.g.,][]{lodato06,  wise08, regan09,
johnson11, agarwal12, dijkstra14}
  are a more plausible scenario than the  direct collapse black holes formed after
galaxy major mergers, as these events are only possible during mergers of much
more 
massive galaxies \citep{mayer10,bonoli14,mayer14}.

\begin{figure*}
\begin{center}
        \includegraphics[width=0.8\textwidth]{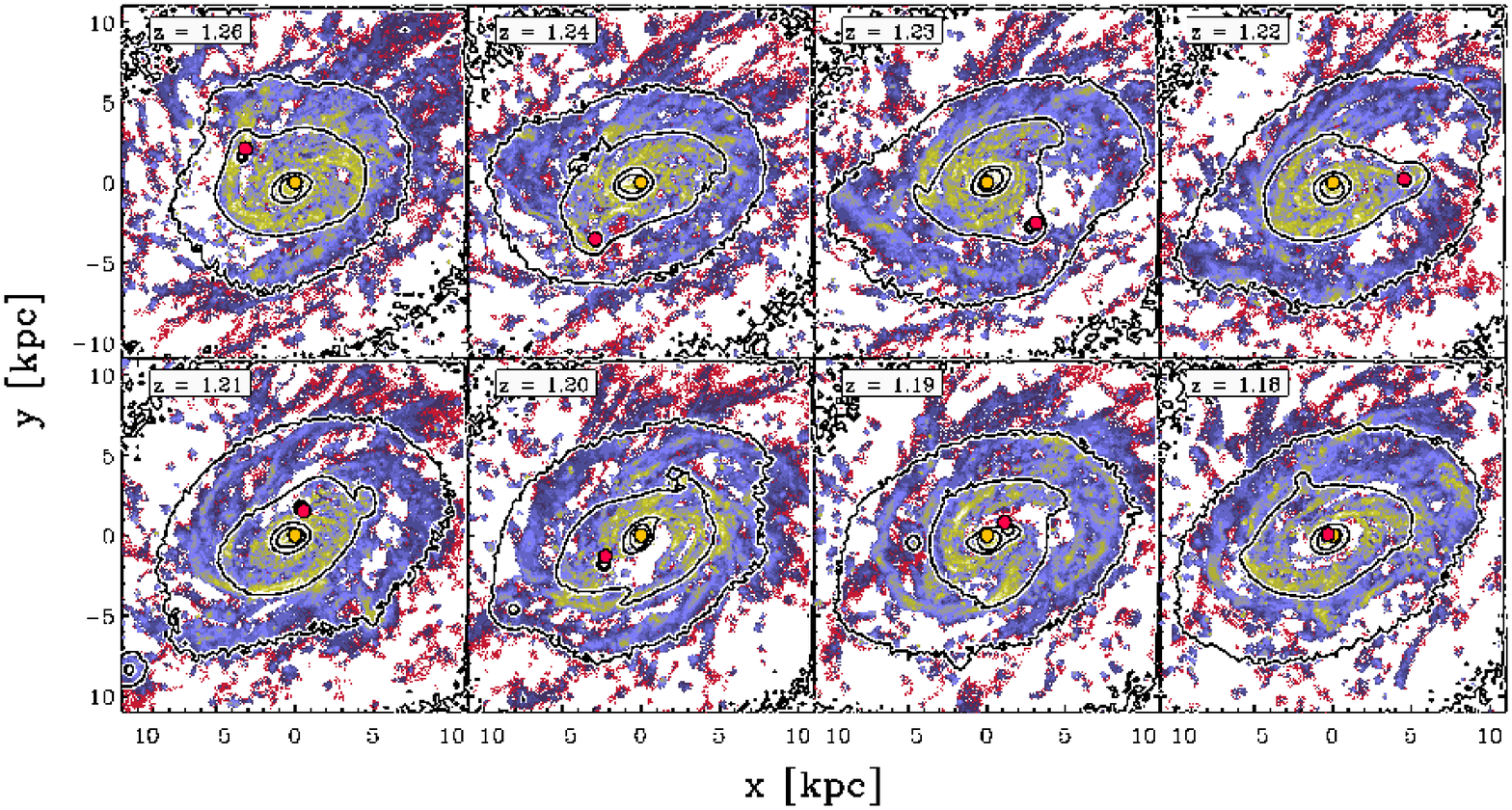}
        \includegraphics[width=0.1\textwidth]{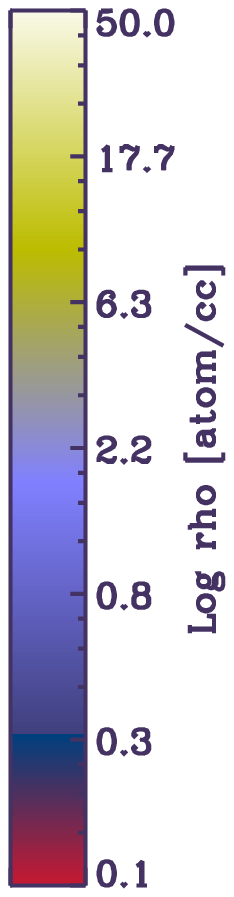}
        \caption{Time sequence of the last black hole merger in the simulation.
Black holes are represented by the yellow and red bullets.
The gas map is color-coded according to the particle density. In black are overplotted stellar isodensity contours.}
        \label{fig:last_merger_map}
\end{center}
\end{figure*}

\subsection{A discussion on the Bondi-Hoyle-Lyttleton prescription}
\label{sec:discussion_bondi}

As introduced in the methodology section, in this paper we have used a specific
model for black hole accretion, the
Bondi-Hoyle-Littleton model, which is simple and widely used
in the literature of cosmological simulations. At the same time, though,  it is known
to be often inaccurate for flows in the nuclei of
 galactic disks, which are neither spherical and hydrodynamical, but are rather
governed by the effects of gravitational torques in
 the redistribution of angular momentum and kinetic energy
\citep[e.g.,][]{debattista06,hopkins10, hopkins11}. In particular,
 \citet{hopkins11} have proposed a different sub-grid model for accretion onto
massive black holes in simulations that is
based on the analysis of angular momentum transport in galactic disks of stars
and gas, where the self-gravitating perturbation
destabilizing the axisymmetric gas flow is assumed to be coming from the stellar
potential only. The latter is a good approximation
in our case since star formation is vigorous in the nucleus at high redshift
(see Figure \ref{fig:sfr_comp}), leading to stellar-to-gaseous mass ratio larger than
$10$ at distances below $300 \pc$ from the center already at redshift $z \sim 4$ (see
Figure \ref{fig:Mass_enclosed_ph}). \citet{hopkins11} have shown that their
proposed new sub-grid model
captures very closely
the mass inflow rates at small scales occurring in very high resolution
simulations, which resolve down to $\sim \pc$ scales, hence close
to the boundary of the accretion disc, while the Bondi model can overestimate by
even more than an order of magnitude the inflow
rate occurring in simulations at such scales. Since we only resolve gravity to
about $100 \pc$ scales, the size of our gravitational softening length, we can
 apply their sub-grid model at such scales. According to this model, the mass inflow
rate $\dot M$ onto the black hole is $\sim a M_{\rm gas}/T_{\rm orb}$, where
$M_{\rm gas}$
is the gas mass at the smallest scale, $T_{\rm orb}$ is the
orbital time at the same distance and $a$ is a constant related to the amplitude
of gravitational torques, which is
maximum in major mergers ($a=1$) and can be as small as $a=0.01$ in weakly
self-gravitating disk. We can apply the model to the
high-redshift  phase of black hole growth ($z = 2-4$) which is the one yielding,
on average,
the highest accretion rates (see Figure \ref{fig:BHmdot_sfr}).
Assuming a $a=0.01-0.1$, since most of the time our galaxy does not undergo significant mergers,
 and taking $M_{\rm gas}$ a few times $10^7 \Msun$ (corresponding to the
values in the original Eris simulation shown in Figure
\ref{fig:Mass_enclosed_ph}) and $T_{\rm orb} \sim 10^7 \yr$,  a typical value
for the inner few hundred parsecs, we obtain $\dot M \sim 10^{-2} - 10^{-1}
M_{\odot} \yr$. Those values for the accretion rate are higher than the accretion
rates measured in the \erisbh simulation,
despite it adopts the Bondi model. This is reassuring, since feedback is
expected to decrease the accretion rates further, yet there is no evidence that
the Bondi model employed is boosting severely
the accretion rates.
This suggests that in our case the accretion rates are not strongly dependent on
the accretion model adopted,  but they are rather
bound to be low because they arise from  the weak gas inflows occurring at the {\it
resolved} scales, from few $\kpc$  to a few softening lengths. This is line
with the notion that we are modeling a fairly quiescent halo that does not
undergo many prominent mergers, and this seems to naturally lead to the formation
of a late-type spiral with a pseudobulge.
It also confirms that our general key conclusion that massive black holes in
late-type spirals grow little by accretion
should be quite general.

\begin{figure}
\begin{center}
        \includegraphics[width=0.49\textwidth]{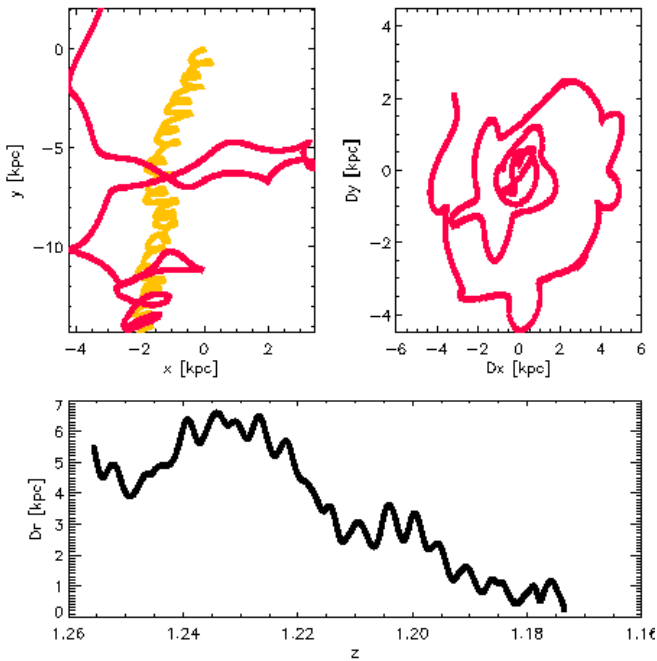}
        \caption{Details of the spatial evolution of the two black holes shown
in the maps of Figure \ref{fig:last_merger_map}. Upper left panel: evolution of
the $x-y$
position of the two black holes. Top right panel: $x-y$ projection of the orbit
of the secondary black hole around the primary. Lower panel: redshift evolution
of the separation between the two black holes.   }
        \label{fig:last_merger_separation}
\end{center}
\end{figure}

\subsection{A close look to black hole mergers} \label{sec:BHmergers}

As previously discussed, mergers with black holes from satellite galaxies
contribute significantly to the mass-growth
budget of the black hole hosted by the central, most massive galaxy of the simulation. The
 resolution of the simulation allows us to follow the evolution of black
hole binaries down to few hundreds of \pc. The maps in Figure
\ref{fig:last_merger_map} show the evolution of the last black hole merger in
the simulation, occurring at $z \sim 1$, with  gas particles colored
according to their density and black lines indicating stellar isodensity contours. At
the beginning of the sequence, the stellar core
 hosting the satellite black hole is still visible in the isodensity contours. 
When the remnant of the satellite galaxy is finally disrupted, the secondary black hole is dragged to
the center of the primary galaxy to eventually merge with the more massive black
hole. This sequence spans $\sim 250 \Myr$ in the cosmology assumed here. 

Figure \ref{fig:last_merger_separation} shows quantitatively the position,
separation and orbit evolution of the two black holes from
several $\kpc$ scales down to the softening scale of the simulation. The upper
left panel shows the $x-y$  position of the two black holes in the same time
interval shown in Figure \ref{fig:last_merger_map}, where the positions have
been centered
at the location of the primary black hole at the time we start tracking the
 binary system. While the central black hole (orange curve) wiggles around its
original position, the satellite black hole shows a typical spiraling pattern 
until it finally joins the central black hole. The top-right panel shows instead
a $x-y$ projection of the orbit of the satellite black hole around the primary
one and, finally, the bottom panel of the same figure shows
the redshift evolution of the separation between the two black holes. In about
$250 \Myr$ the black holes go from a distance of several $\kpc$ to  a separation
smaller than the spatial resolution of the simulation, at which point they
 are assumed to merge.  Such short timescale for the evolution of the black
hole binary system is consistent with the results of high-resolution merger
simulations of, e.g.,  \citep{mayer07,chapon13}. 

 However, we  also note that detailed numerical simulations of massive black hole mergers
in circumnuclear disks have shown that the formation of a tight binary below
our resolution
scale of $120 \pc$ can be
delayed to up to $10^8 \yr$  in the favourable case of major mergers \citep{roskar15},
and to even longer timescales (up to a $\Gyr$) in the case of minor mergers, due to
a variety of processes such as ram pressure stripping of the secondary galaxy
core
surrounding the secondary black hole \citep{callegari09,callegari11}. In any
case, all black hole mergers in our simulation occur at very early times, when
dynamical timescales are still very short, possibly explaining that  neglecting to capture the small scale dynamics is not an issue.
We finally note that, with our resolution, we encounter no problems in
following the black hole orbital decay directly down to the scale of the softening
length, namely without having to impose an artificial drag force \citep{tremmel15}.

\begin{figure}
\begin{center}
        \includegraphics[width=0.49\textwidth]{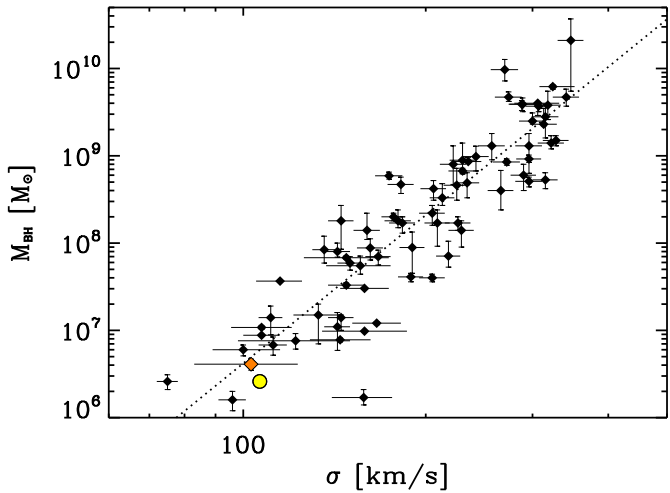}
        \caption{$\MBH-\sigma$ plane: the black symbols are observational points
 from the compilation of \citet{mcconnell13} (best fit shown by the dotted black
line), the orange diamond is the location of the Milky Way in the plane while
the yellow circle indicates the location of \erisbh.}  
        \label{fig:MBH_sigma}
\end{center}
\end{figure}

\begin{figure*}
\begin{center}
        \includegraphics[width=0.49\textwidth]{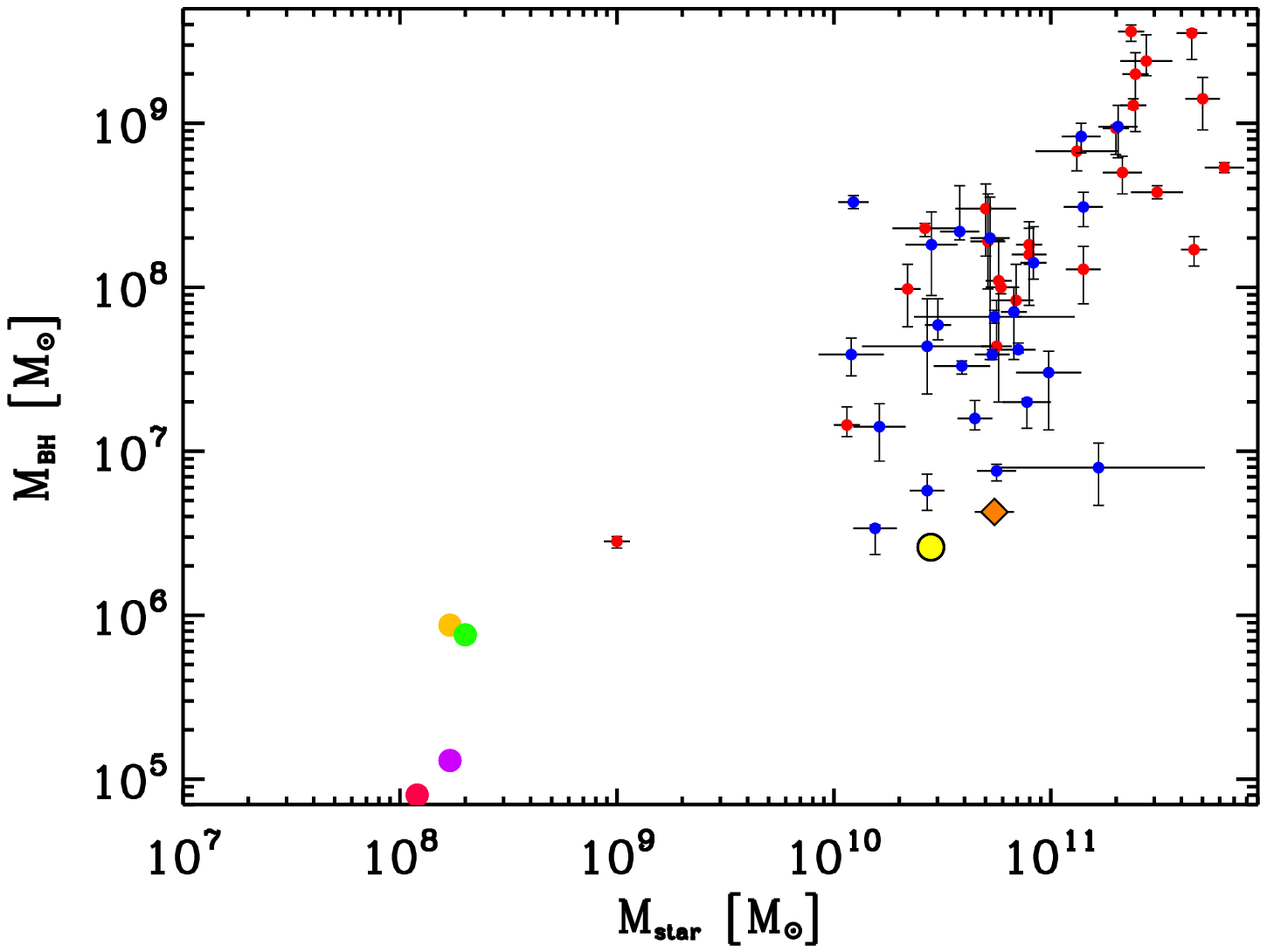}
        \includegraphics[width=0.49\textwidth]{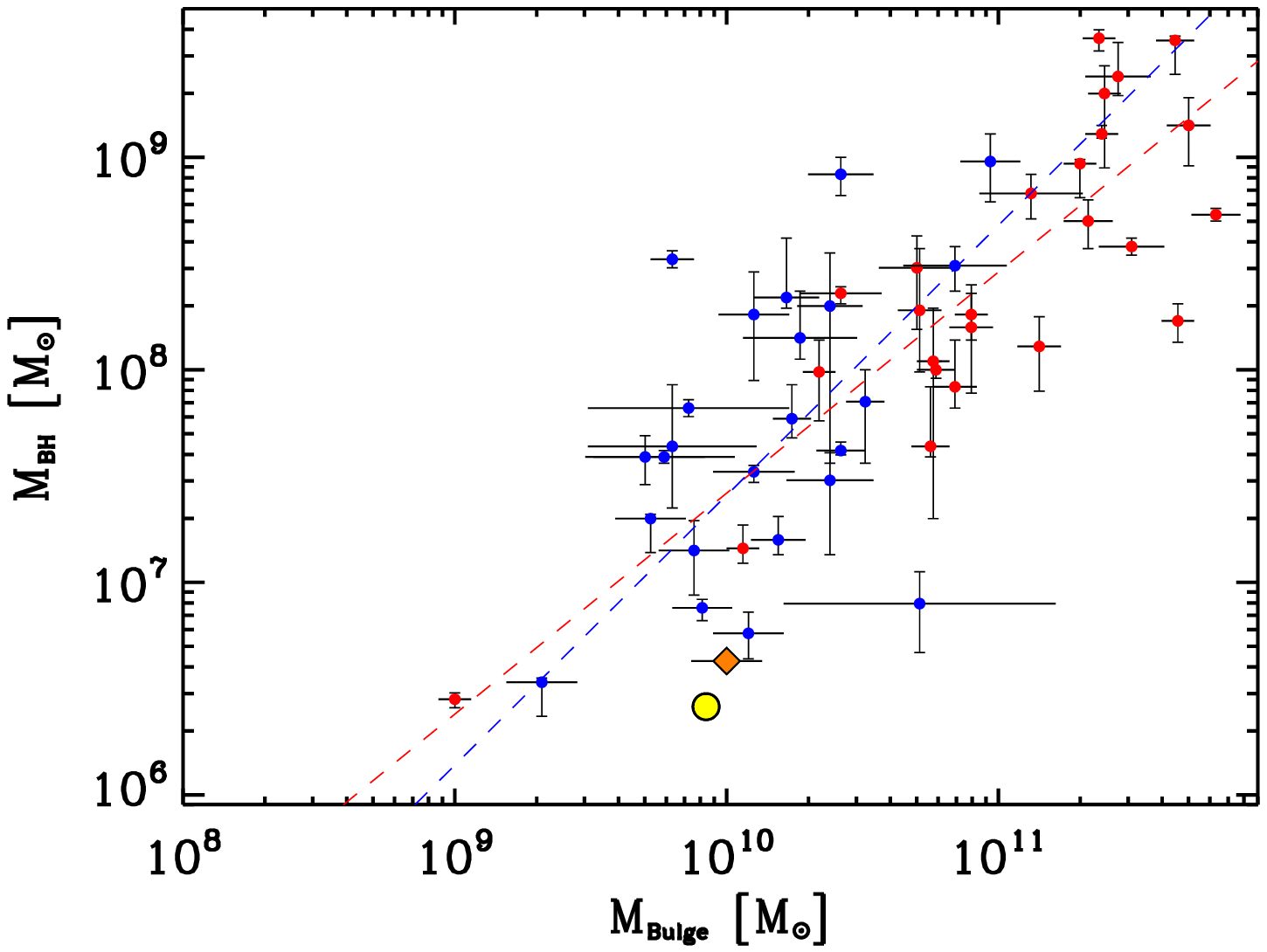}
        \caption{$\MBH-M_{\rm Star}$ and $\MBH-\MBulge$ relations. The red and
blue symbols are from a compilation of \citet{erwin12} for early type and late
type galaxies, respectively (with fits for the $\MBH-\MBulge$ relation shown by
the dashed lines). The orange diamond is the location of the Milky Way
in the plane as reported by  \citet {erwin12}. The yellow circle indicates the
location of \erisbh. The
colored bullets on the left of the plot show the location of the seeds,
color-coded as in Figure  \ref{fig:map_seed} }  
        \label{fig:MBH_star}
\end{center}
\end{figure*}

\subsection{Black hole-galaxy scaling relations}  \label{sec:scalings}

Since their discovery more than 20 years ago
\citep[e.g.,][]{magorrian98,merritt01,tremaine02}, the tight relations
between black hole mass and various properties of the host galaxies have been
interpreted as an evidence of some type of co-evolution between massive black
holes and their hosts. Those relations have been first defined for massive
bulges or elliptical galaxies, and has not been clear whether they hold when going to
low mass galaxies or galaxies with different morphologies. Recent studies
\citep[e.g.,][]{greene08, hu08, kormendy11, jiang11, mathur12,graham15} have found that black holes in
pseudobulges do not correlate in the same way, if at all, with galaxy properties
as black holes in classical bulges. 
Studying active low-mass galaxies, \citet{greene08} found, for example, that
black holes in pseudobulges sit on the extrapolation
 to lower masses of the $\MBH-\sigma$ relation, while they lie almost an order
of magnitude below  the
$\MBH-\MBulge$ relation. The different behaviour
 for classical bulges and pseudobulges seems to be due to
the different relation between stellar mass and velocity dispersion that they
follow   \citep{gadotti09}. \citet{kormendy11} studied systems with
dynamically-estimated black hole masses, and found evidence that no correlation
actually exists between black hole masses and
pseudobulges. Those authors argue that this is likely due to the different history
  of black holes evolving in pseudobulges and the ones in classical bulges: black
hole growth in pseudobulges is likely  driven
by small-scales and stochastic events and is typically highly sub-Eddington,
in contrast with the growth of the most massive black holes, which is driven by  large scale dramatic
events, such as the violent mergers, that are able to both efficiently feed
black holes and to generate classical bulges and elliptical galaxies.
The bulge of \erisbh  has the properties of a
pseudobulge rather than a classical bulge (see Section \ref{sec:star}),
similarly to the bulge of Eris, whose evolution has been studied in a dedicated
paper by \citet{guedes13}. We can thus directly see whether our black hole has
properties consistent with the ones of black holes in observed pseudobulges.
 
In Figure \ref{fig:MBH_sigma} is shown where the \erisbh central black hole sits
in the $\MBH-\sigma$ relation (yellow bullet), compared with the observational
 data points  compiled by \citet{mcconnell13}. We estimated the  velocity
dispersion for \erisbh using the 
$1/\sqrt{3}$ of the $3-D$ velocity dispersion within $2 \kpc$ from the center. With this assumption, we obtain a value for $\sigma$
of $109 \kms$. Our calculation for the velocity dispersion gives values that are
quite comparable to observational long-slit measurements of face-on nearby
galaxies \citep{bellovary14}.
 The obtained value of  $109 \kms$ is also within the errors on  the velocity dispersion  of
$103 (\pm 20) \kms$
for the Milky Way
given in the compilation of \citet{mcconnell13}.
We see that  \erisbh sits slightly below 
the relation, as its black hole mass is about $2/3$ the mass of the Milky
Way central black hole, assuming that SgrA$*$, as in \citet{mcconnell13}, is 
$4.1 \times 10^6 \Msun$ \citep[from:][]{ghez08,gillessen09}.

Figure \ref{fig:MBH_star} shows instead the scaling relations between black hole
mass and stellar mass and between black hole mass and the bulge component of
galaxies. Here we use as reference observational points the compilation of
\citet{erwin12} (red and blue symbols, representing the location of early and
late type galaxies respectively). As in the previous figure, the orange diamond
represent the location of the Milky Way and the yellow bullet the location of \erisbh.
 In Section \ref{sec:star} we perform a bulge and disc decomposition by fitting a
double-S\'ersic profile to the surface density
profile of our simulated galaxy: integrating the best-fit parameters of the S\'ersic
profiles, we obtain estimates for the bulge and disk masses (see Table
\ref{table:info_global}).  The stellar mass used here  is the sum of the bulge and
disc derived from the decomposition. \erisbh  sits well below the relation, even though
the black hole seeds start off close to the extrapolation of the relation at lower masses: this is because the black hole
accretion rates are much lower than what would be required to keep up with the
star formation rate in the central region (see Figure \ref{fig:BHmdot_sfr}). 

Having only one simulation, we can not give quantitative predictions on whether
black holes in pseudobulges follow a different relation or do not follow any
relation at all with the properties of the host. But we note that the
consistency with the $\MBH-\sigma$ relation on one side, and the deviation
 from  the $\MBH-\MBulge$ on the other side, are consistent with what is found
by \citet{greene08}, but is also consistent with  the picture drawn by
\citet{kormendy11}, given the slow Seyfert-like growth of our simulated
black hole.

\begin{figure*}
\begin{center}
        \includegraphics[width=0.49\textwidth]{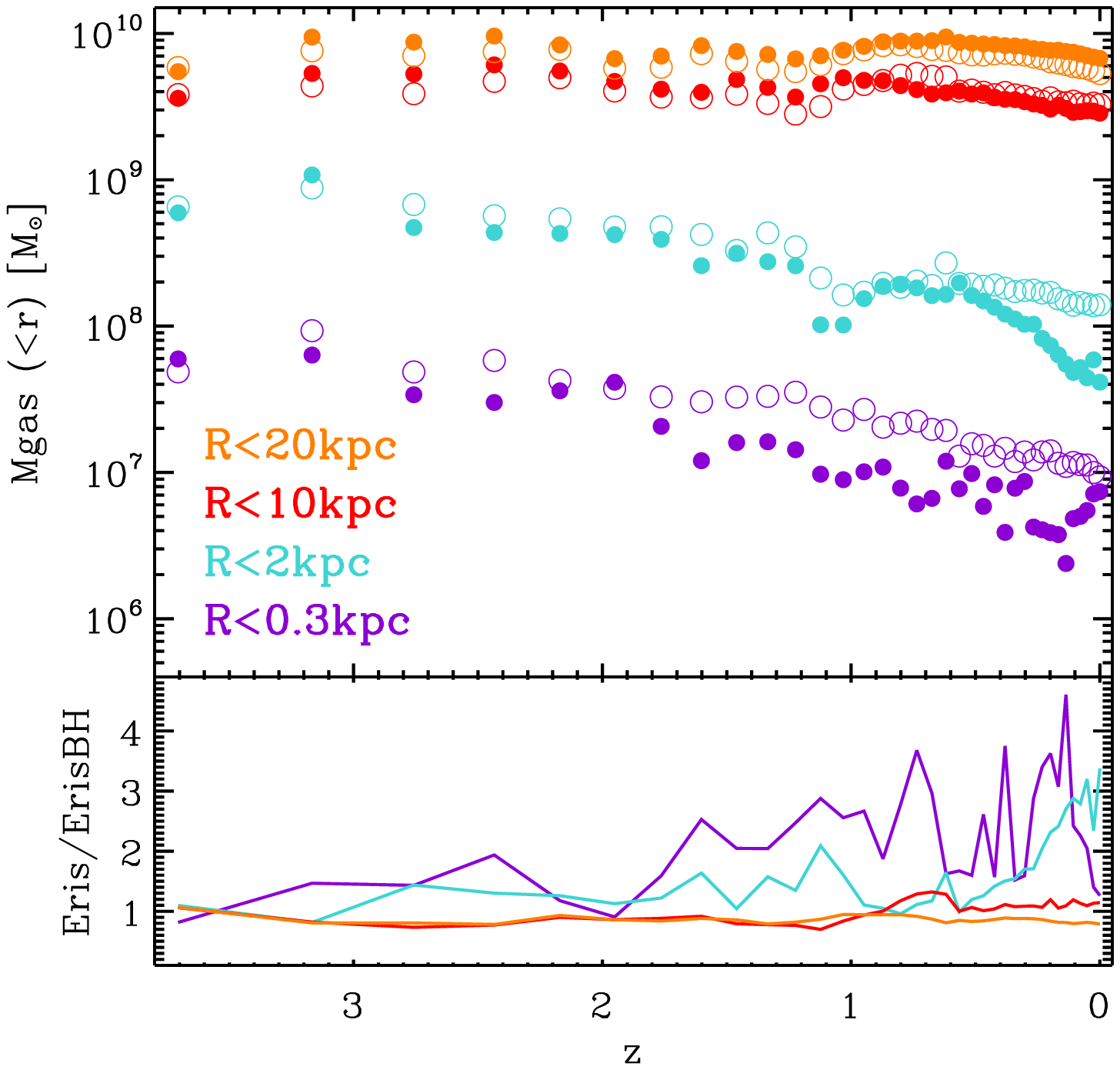}
        \includegraphics[width=0.49\textwidth]{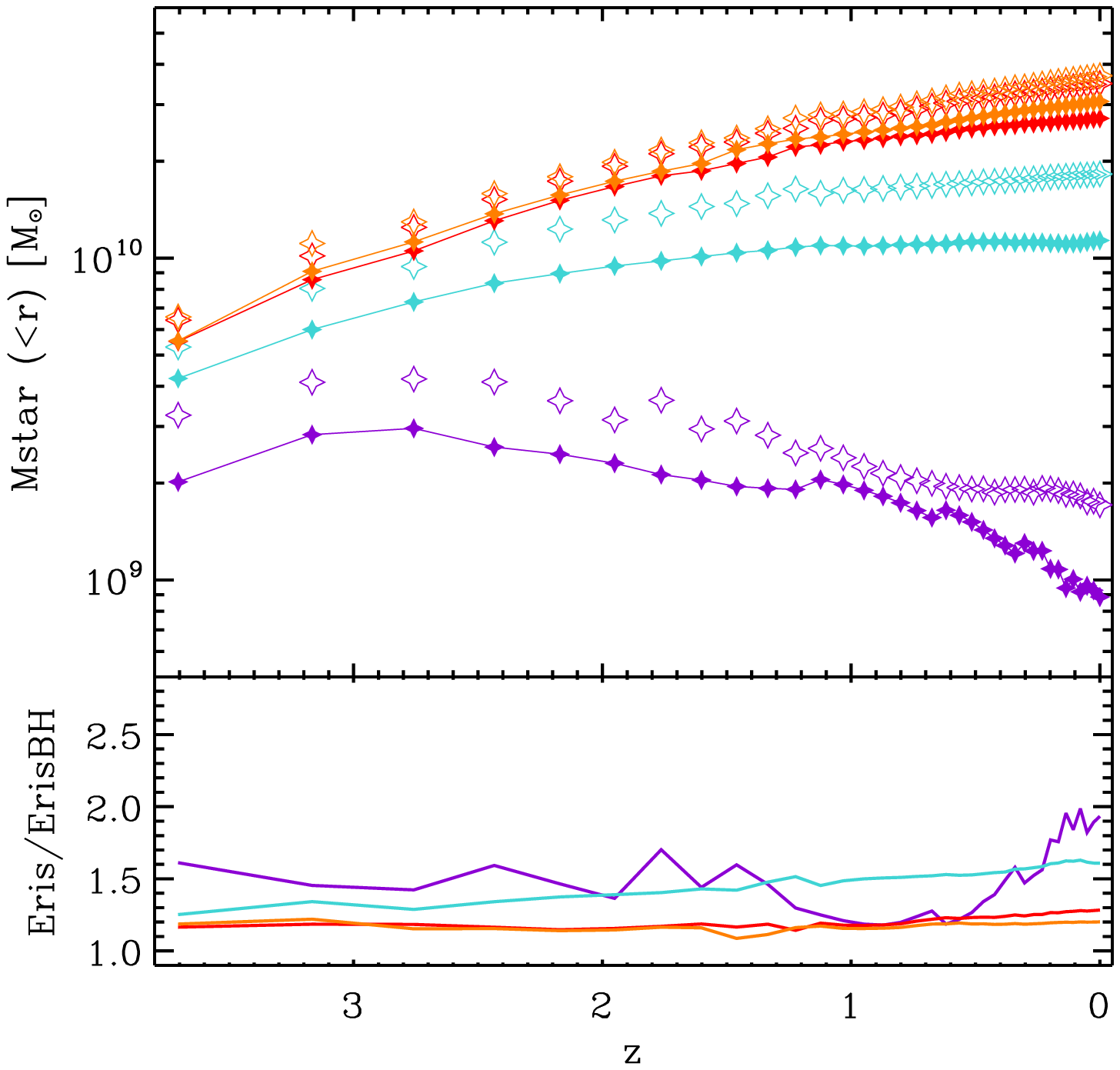}
        \caption{Redshift evolution of gas mass (left panel) and star mass
(right
panel) enclosed within different radii from the galaxy center. The
filled circles are for \erisbh, while the open ones for Eris. The bottom panels
show the ratios of the corresponding enclosed mass between Eris and \erisbh.}
        \label{fig:Mass_enclosed_ph}
\end{center}
\end{figure*}

\begin{figure*}
\begin{center}
        \includegraphics[width=0.49\textwidth]{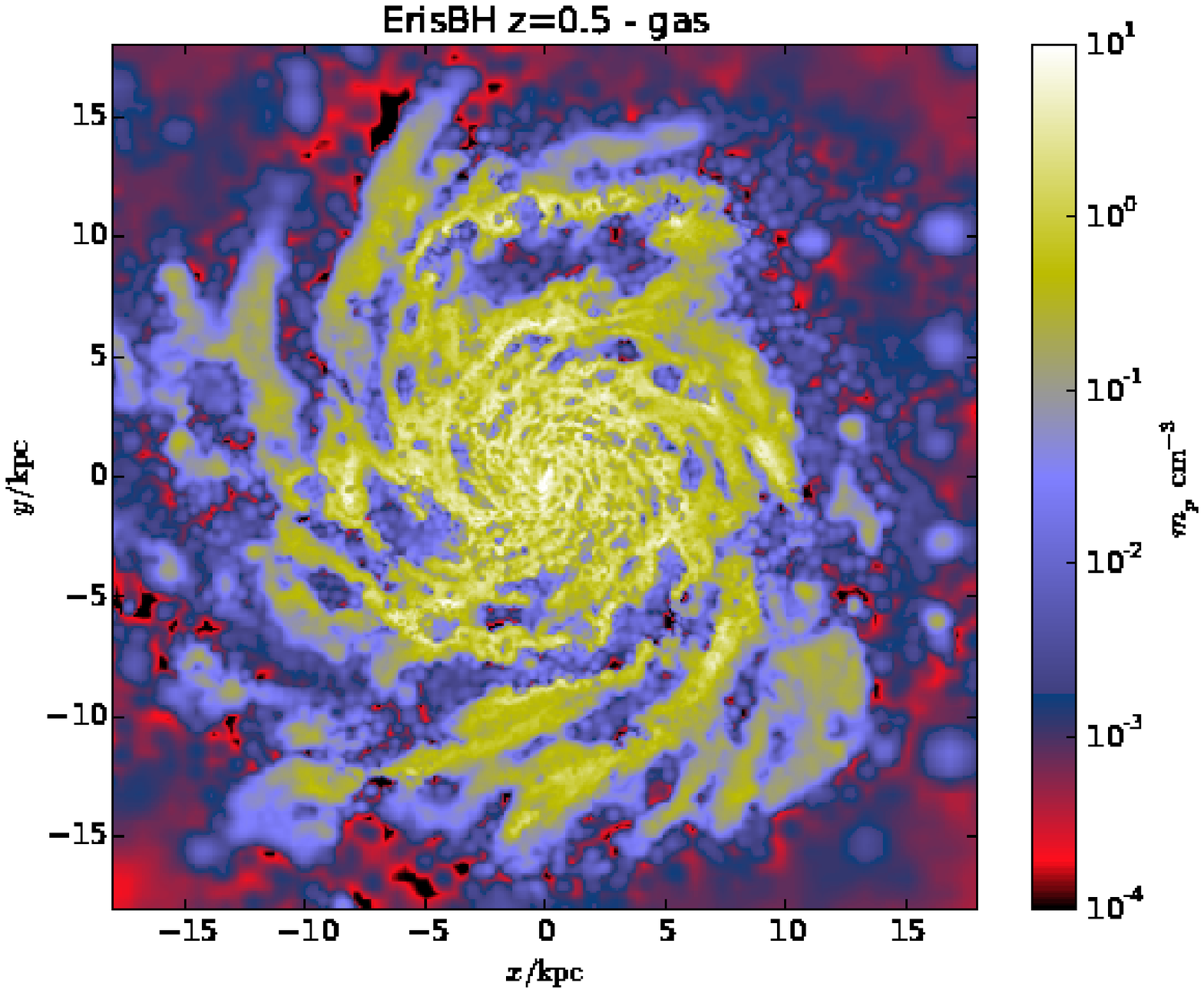}
        \includegraphics[width=0.49\textwidth]{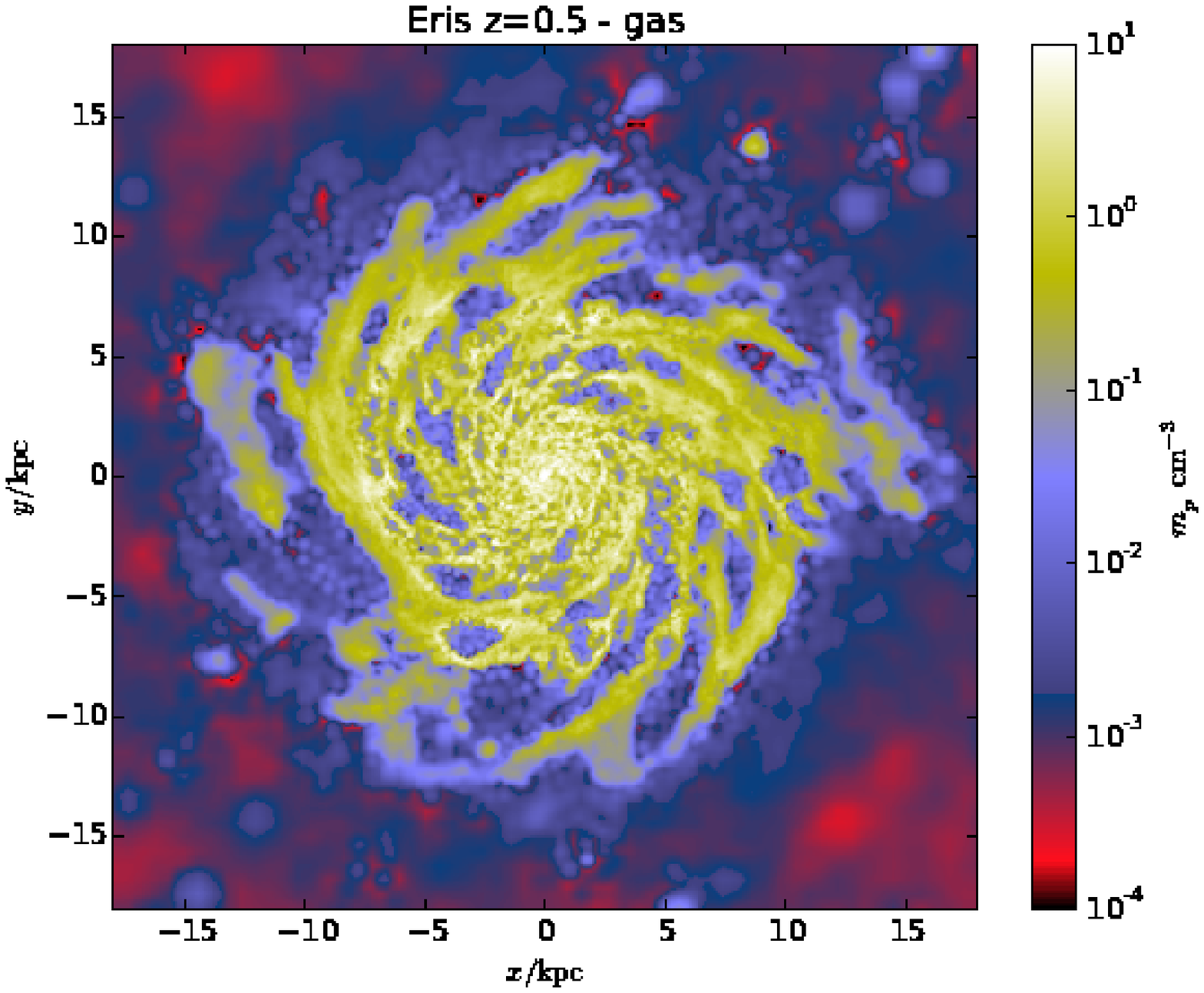}
        \includegraphics[width=0.49\textwidth]{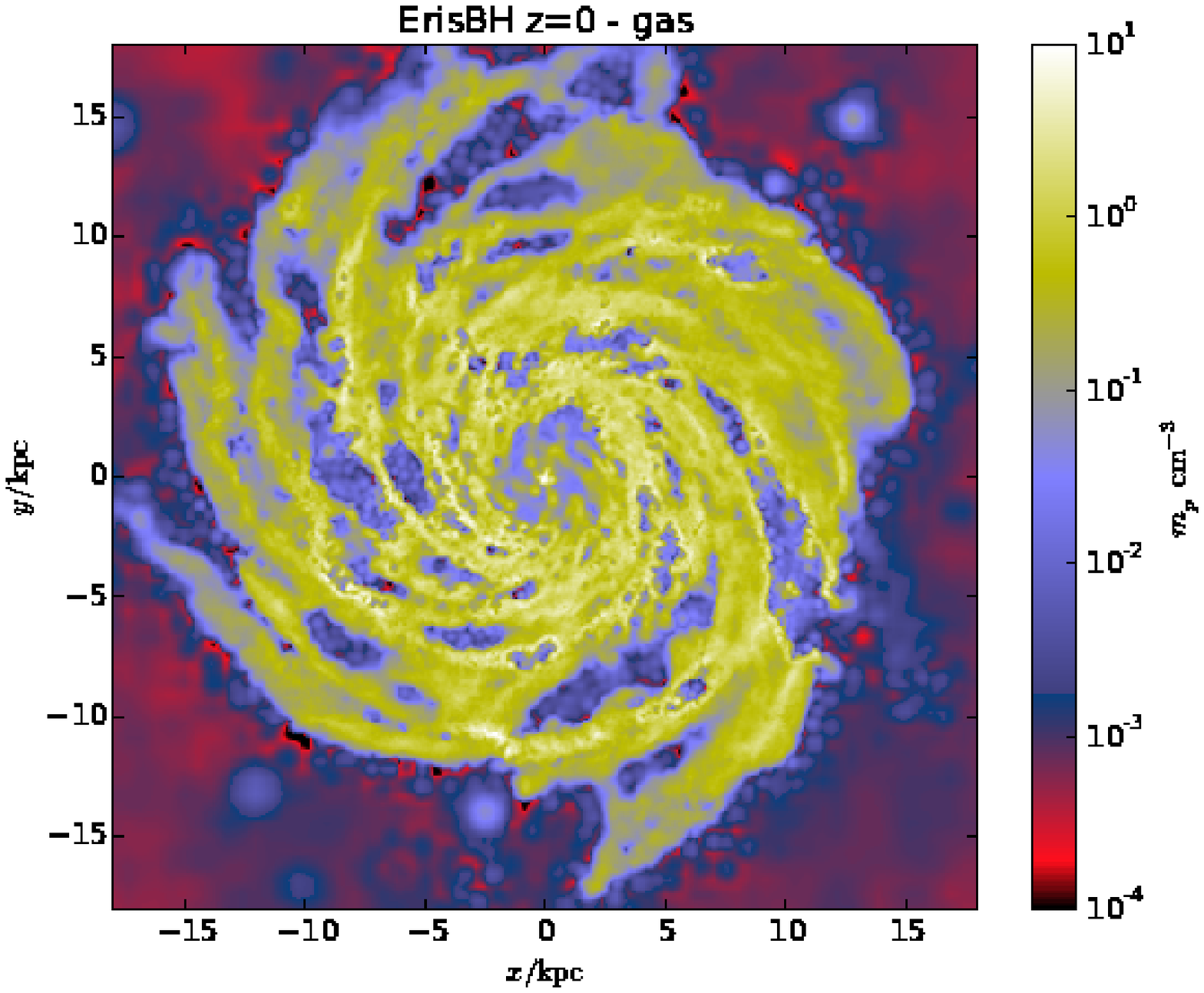}
        \includegraphics[width=0.49\textwidth]{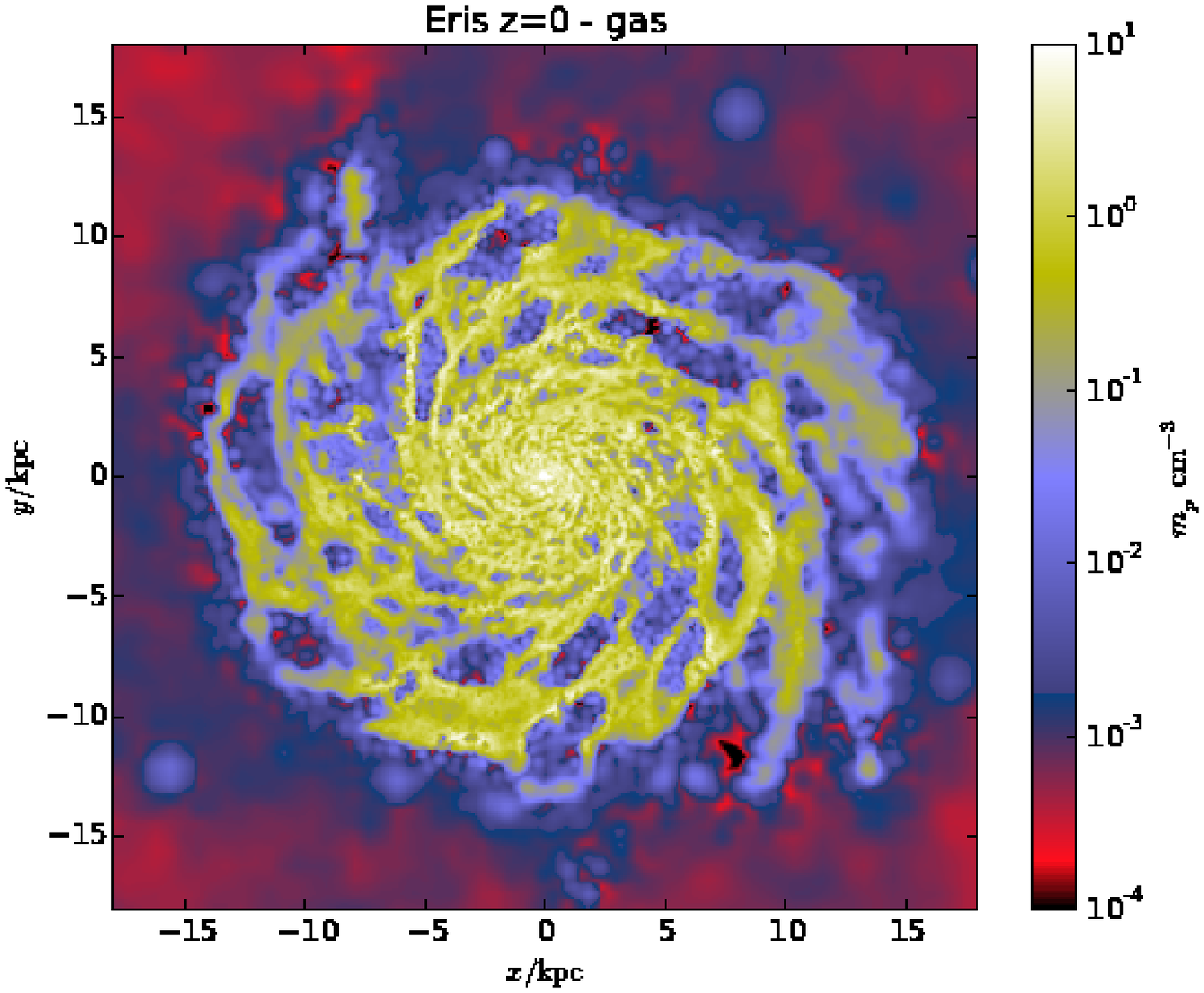}
        \caption{Gas density maps at $z=0.5$ (upper panels) and at $z=0$ (lower
panels) for \erisbh (left) and  Eris (right).}
        \label{fig:Gas_map}
\end{center}
\end{figure*}

\section{Black hole feedback and its effects on the host galaxy} \label{sec:Galaxy}

As described in Section \ref{sec:acc_feedback}, the simulation includes also a
prescription for black hole feedback: a fraction of the energy extracted in the
accretion process couples thermally with the surrounding medium. This standard
prescription has been used extensively in other works, and its effects led to
the idea of a  ``self-regulation'' of black hole growth to explain the origin of the
$\MBH-\sigma$ relation and the ejection of gas from the host galaxy and the
 resulting quenching of star formation \citep[e.g.,][]{diMatteo05}. Given that the accretion
rates are generally quite small for the central black hole of our simulated
galaxy, we expect the effects of feedback to be modest. 
We find, in fact, that the amount of energy released by the black hole is
significantly lower than the one released by supernova explosions (integrating
 from $z \sim 3$ to $z=0$, the total energy input from supernovae, both of type Ia and
II, is more than an order of magnitude higher than the energy input from the
central black hole). Despite not playing any dramatic role in
shaping the host galaxy, AGN feedback seems, however, to be able to keep hot the very central
region of the galaxy, preventing the bulge to grow
  in the same way as in the original Eris.

To isolate the effects of AGN feedback on the \erisbh galaxy, we compare the
properties of its gaseous and  stellar components  directly with the Eris
galaxy, where AGN feedback was ignored in the calculation.

\subsection{Gas component} \label{sec:gas}

In the left panel of Figure
\ref{fig:Mass_enclosed_ph}  we show the
redshift evolution of the total mass in gas of Eris and \erisbh within spheres of
different radii from
the galaxy center. At high redshift (above $z\sim0.5$), the gas content of the
two simulations is very similar, except in the very inner region (purple
curves), where the gas budget of \erisbh is systematically lower than the one of
Eris, a consequence of  the weak feedback from the slowly growing black hole.  

\begin{figure}
\begin{center}
        \includegraphics[width=0.49\textwidth]{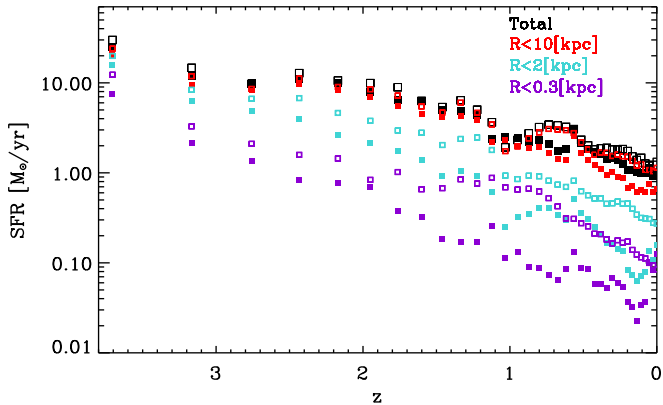}
        \caption{Comparison between the star formation history of \erisbh (filled
squares) and Eris (empty squares). The black symbols show the evolution of the
total sfr ($\Msun/\yr$) in the simulation, while the colored symbols show the sfr within 
 a given radius  from the center of the galaxy, as stated in the legend. }
        \label{fig:sfr_comp}
\end{center}
\end{figure}

While at large radii the total mass in gas remains very similar 
 in the two simulations down to the present time,  in the  inner $2 \kpc$ of
\erisbh (cyan curves) the amount of gas starts decreasing at low redshift 
and, by $z =0$, it is about a third of the one of Eris. The steep decline in gas
content at those scales starts around $z\sim 0.3-0.4$, which is 
 when  the strong bar visible in the galaxy at $z=0$ starts forming (see next
section); as the bar gets stronger, it ``clears'' the center of the galaxy of
its gas content. An exception is the very central region (the inner few hundred parsecs,
purple curves), where the amount of gas slightly increases at those late times;
this raise of gas mass in the center of the galaxy
 results in a slight increase of the black hole accretion rate (see Figure
\ref{fig:BHmdot_sfr}) as well as of the star formation rate (see Figure
\ref{fig:sfr_comp}).

 In Figure \ref{fig:Gas_map} we show density maps at $z=0.5$ (upper panels) and $z=0$
(lower panels) in a region of $18 \kpc$ of radius  around the center of
\erisbh(left column)  and Eris (right column). The angle of projection is chosen
so that the disks appear face-on. 
While, as just discussed, the total amount of gas at large scales is the approximately the same in
both simulations, gas seems to be more diffuse 
 in the \erisbh galaxy, as also its  stellar disk is
larger than the one of Eris (see the next subsection).   But the most striking
difference between the two simulations is probably the  ``empty'' region, of about $2.5 \kpc$
of radius, in the center of  \erisbh at $z=0$. This region has approximately the physical extent of the bar that has formed at $z
\sim 0.3$ and that, as we mentioned above, by $z=0$ has cleared the center of the galaxy.

\begin{figure*}
\begin{center}
        \includegraphics[width=0.49\textwidth]{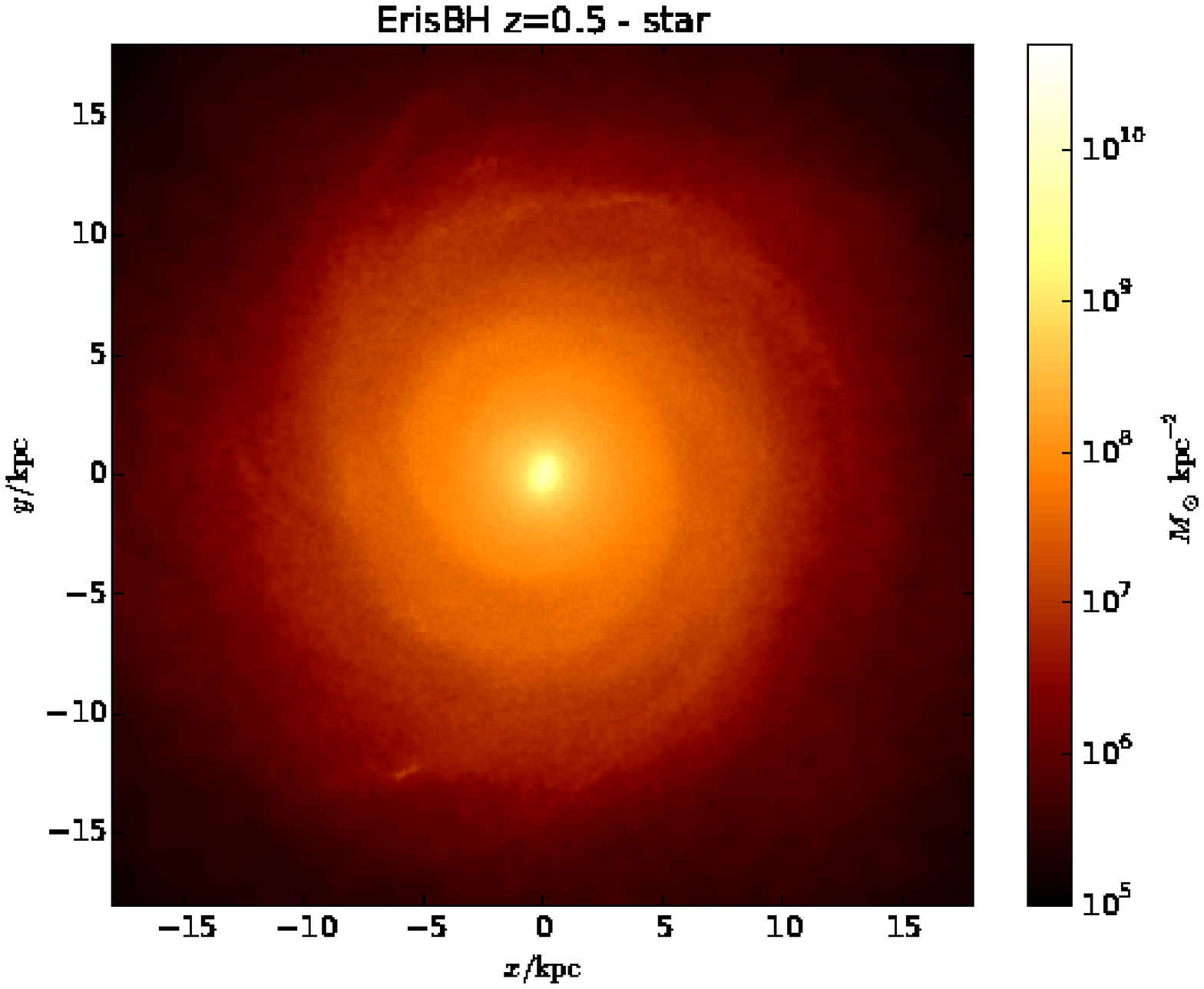}
        \includegraphics[width=0.49\textwidth]{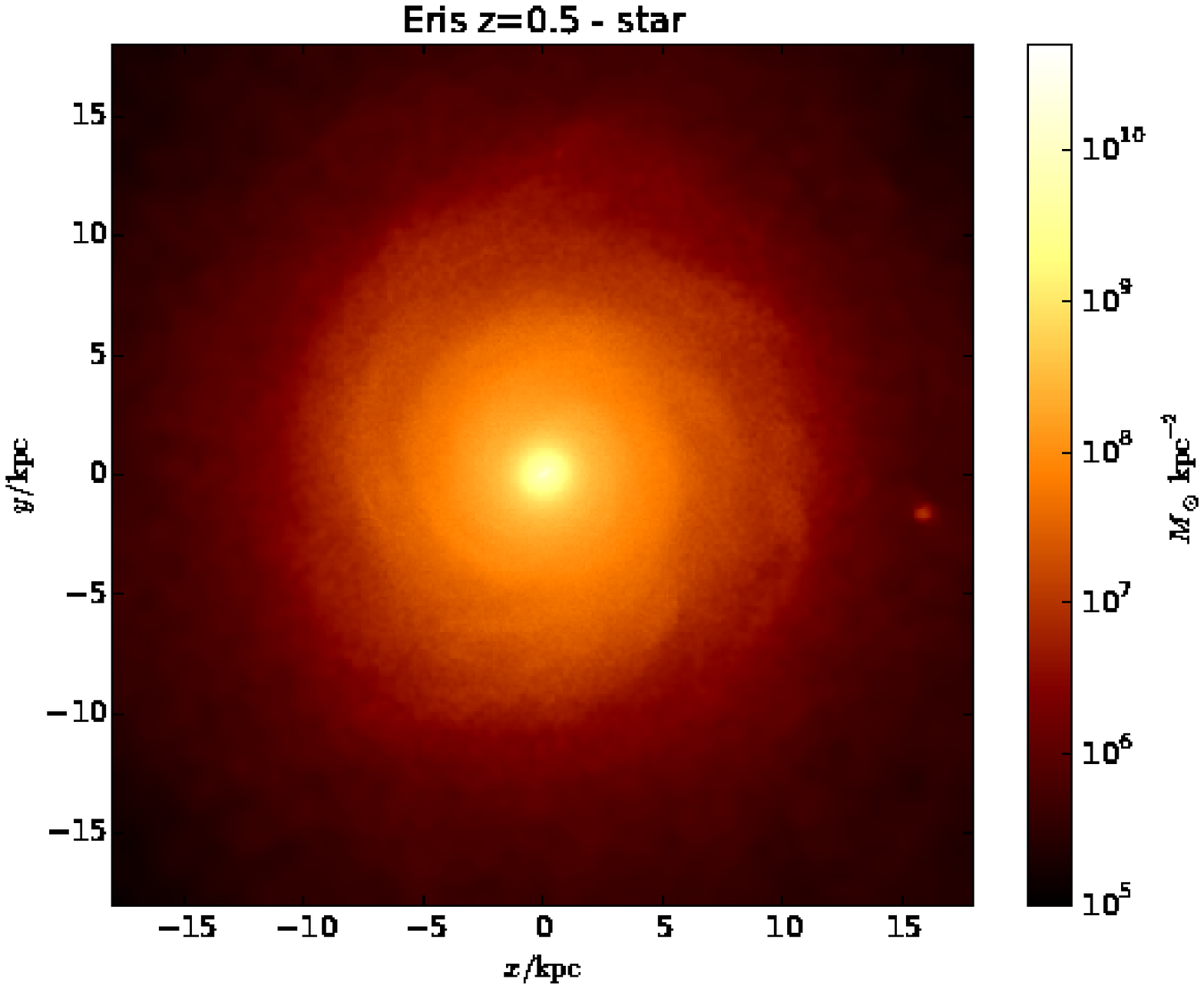}
        \includegraphics[width=0.49\textwidth]{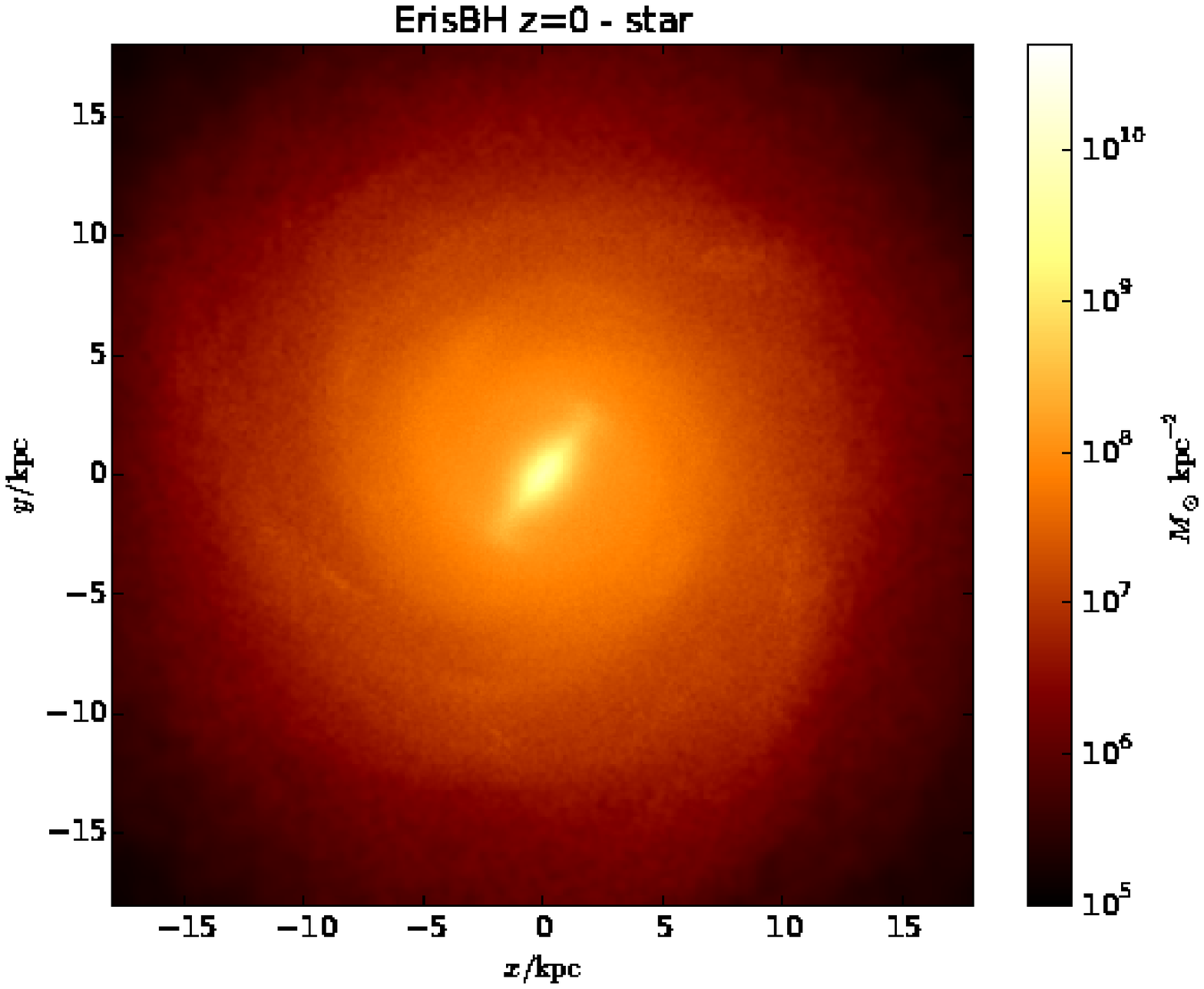}
        \includegraphics[width=0.49\textwidth]{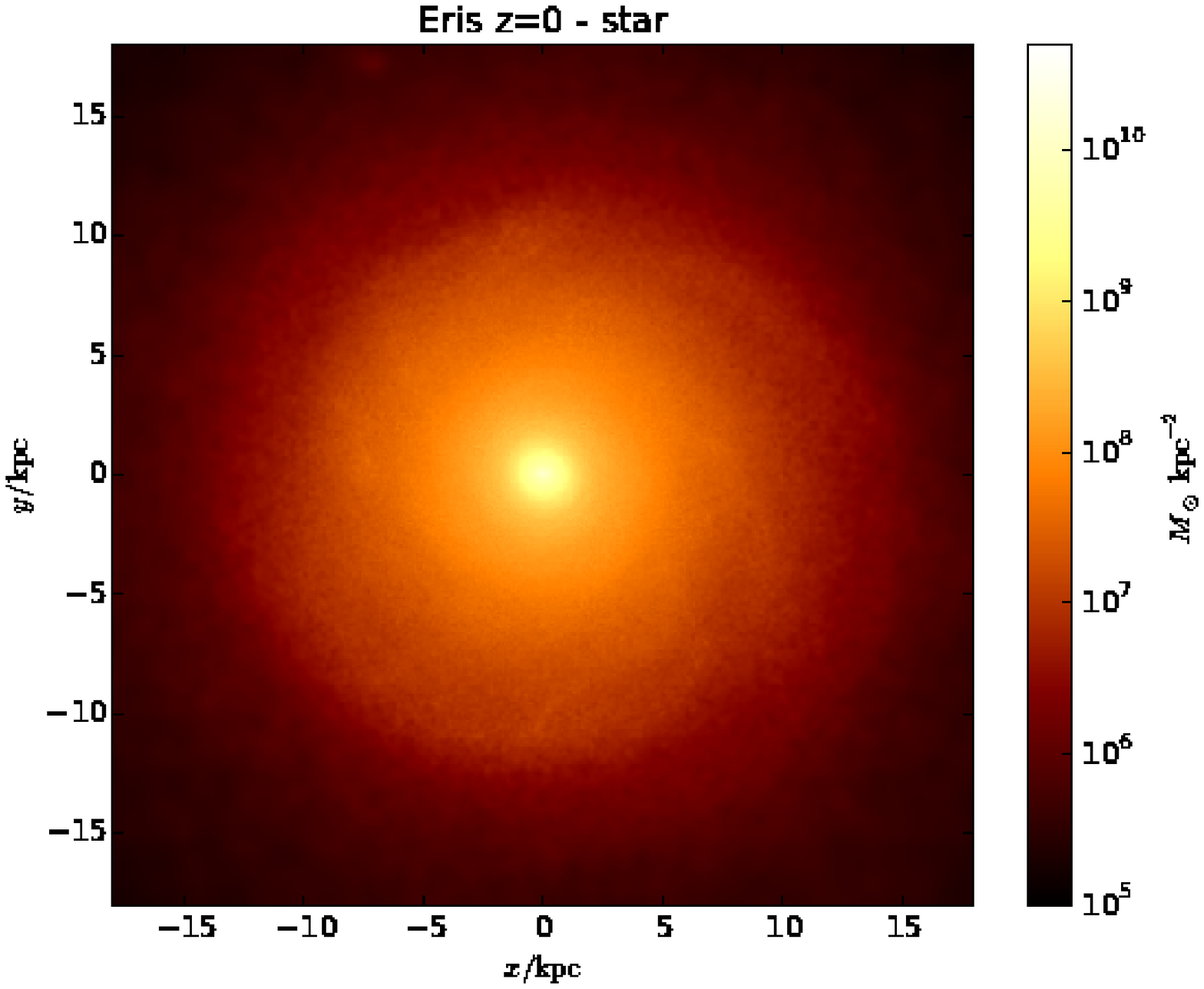}
        \caption{Star density maps at $z=0.5$ (upper panels) and at $z=0$ (lower
panels) for \erisbh (left) and  Eris (right). }
        \label{fig:Star_map}
\end{center}
\end{figure*}

\begin{figure*}
\begin{center}
        \includegraphics[width=0.49\textwidth]{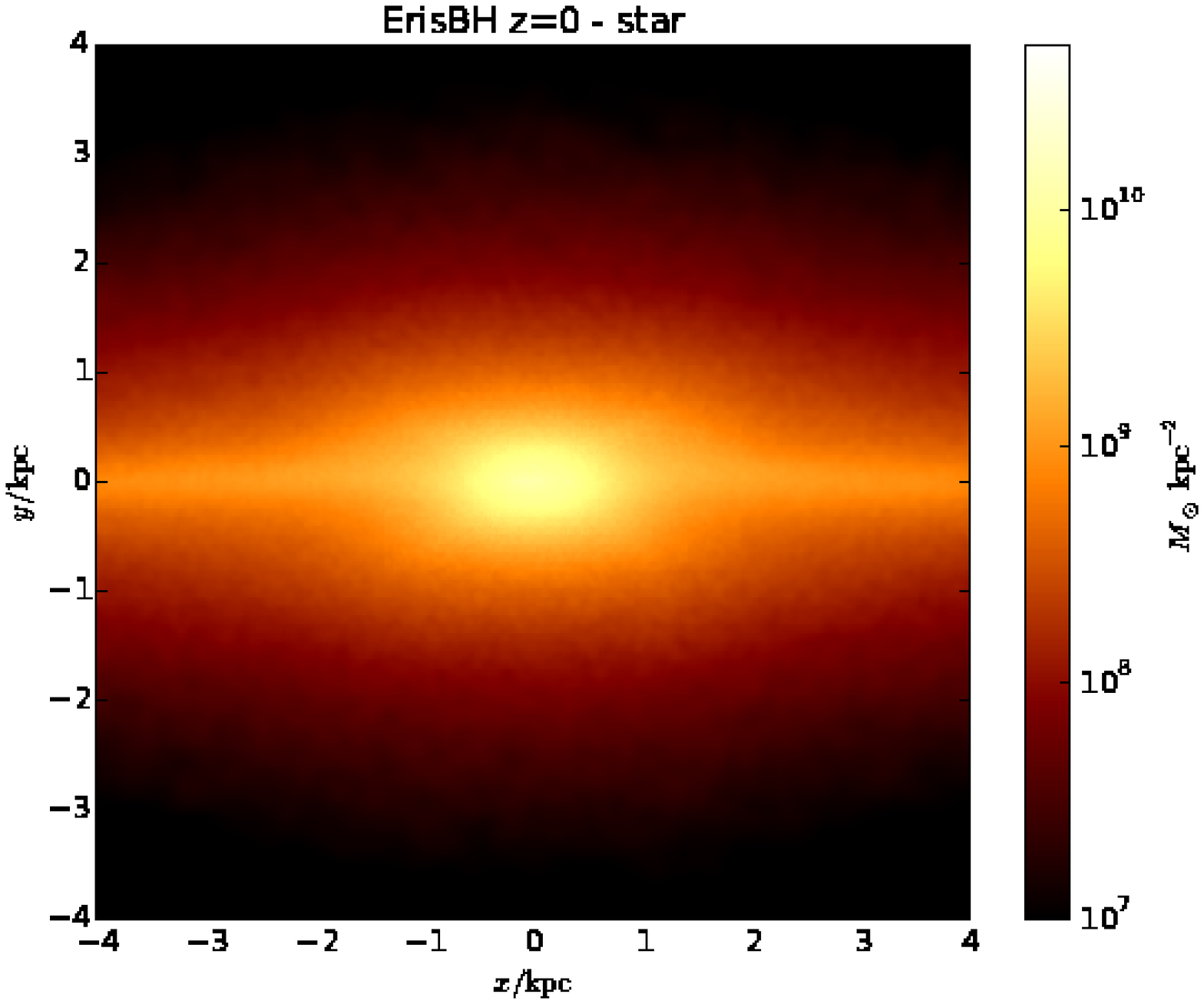}
        \includegraphics[width=0.49\textwidth]{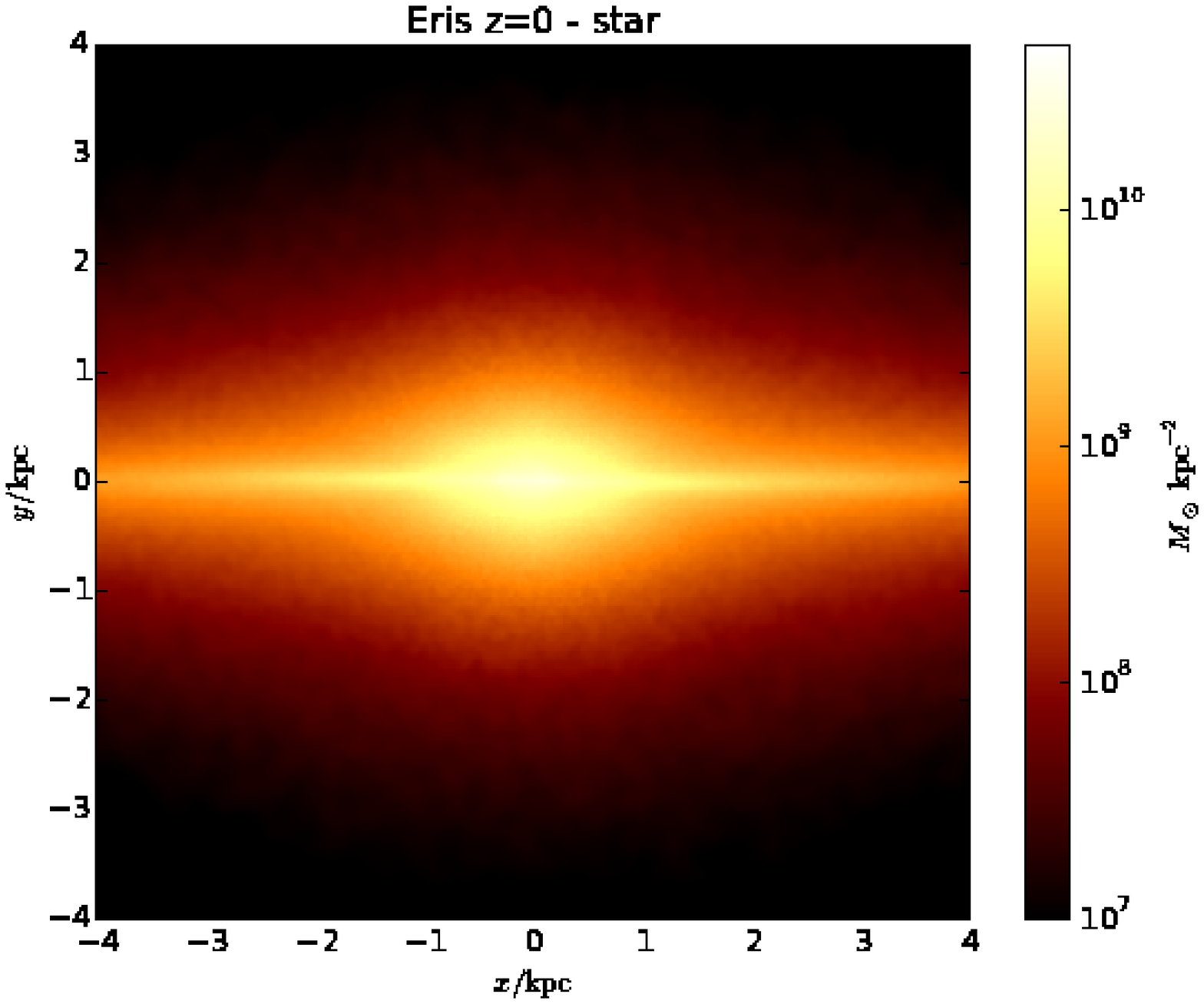}
	\caption{Zoom into the central region of \erisbh (left) and Eris (right) at $z=0$, now viewed with the
disk edge-on.  }
	\label{fig:Zoom_Star}
\end{center}
\end{figure*}

\subsection{Stellar component} \label{sec:star}

Figure \ref{fig:Star_map} shows maps of the stellar density projected face-on, for the same regions
of Figure \ref{fig:Gas_map}. At $z=0.5$ the main differences between the two
simulations are the slightly larger disk and the smaller bulge of \erisbh. 
 It is again at $z=0$ where we see the most remarkable differences between the
two simulations: in \erisbh the 
bulge is not only smaller than the one of Eris, but now it also features a
strong bar, clearly visible in the map, and which has a radius of about $3
\kpc$.  The presence of the bar causes the bulge to have a boxy-peanut
morphology when viewed edge-on, as shown in Figure \ref{fig:Zoom_Star}. 

\begin{figure}
\begin{center}
        \includegraphics[width=0.23\textwidth]{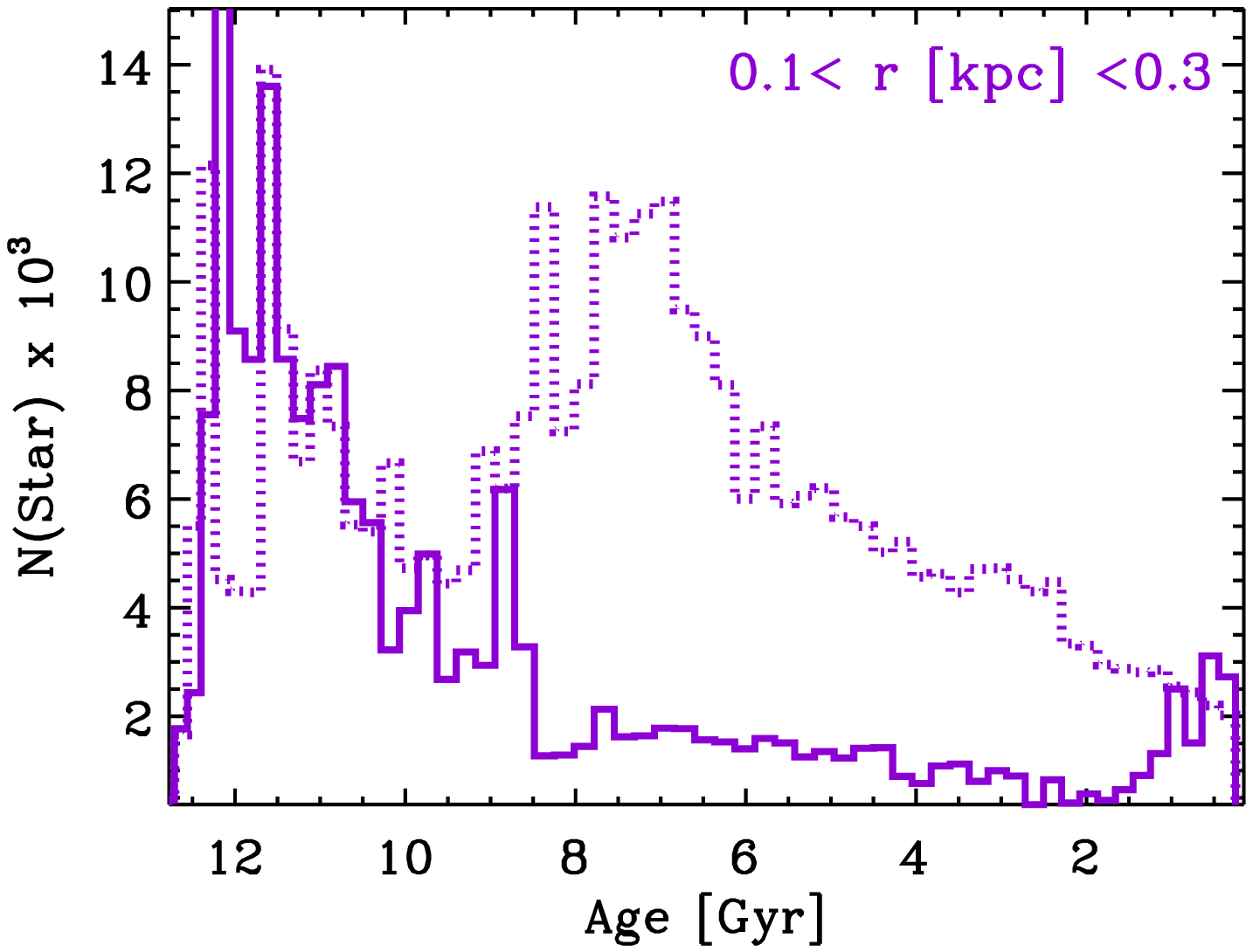}
        \includegraphics[width=0.23\textwidth]{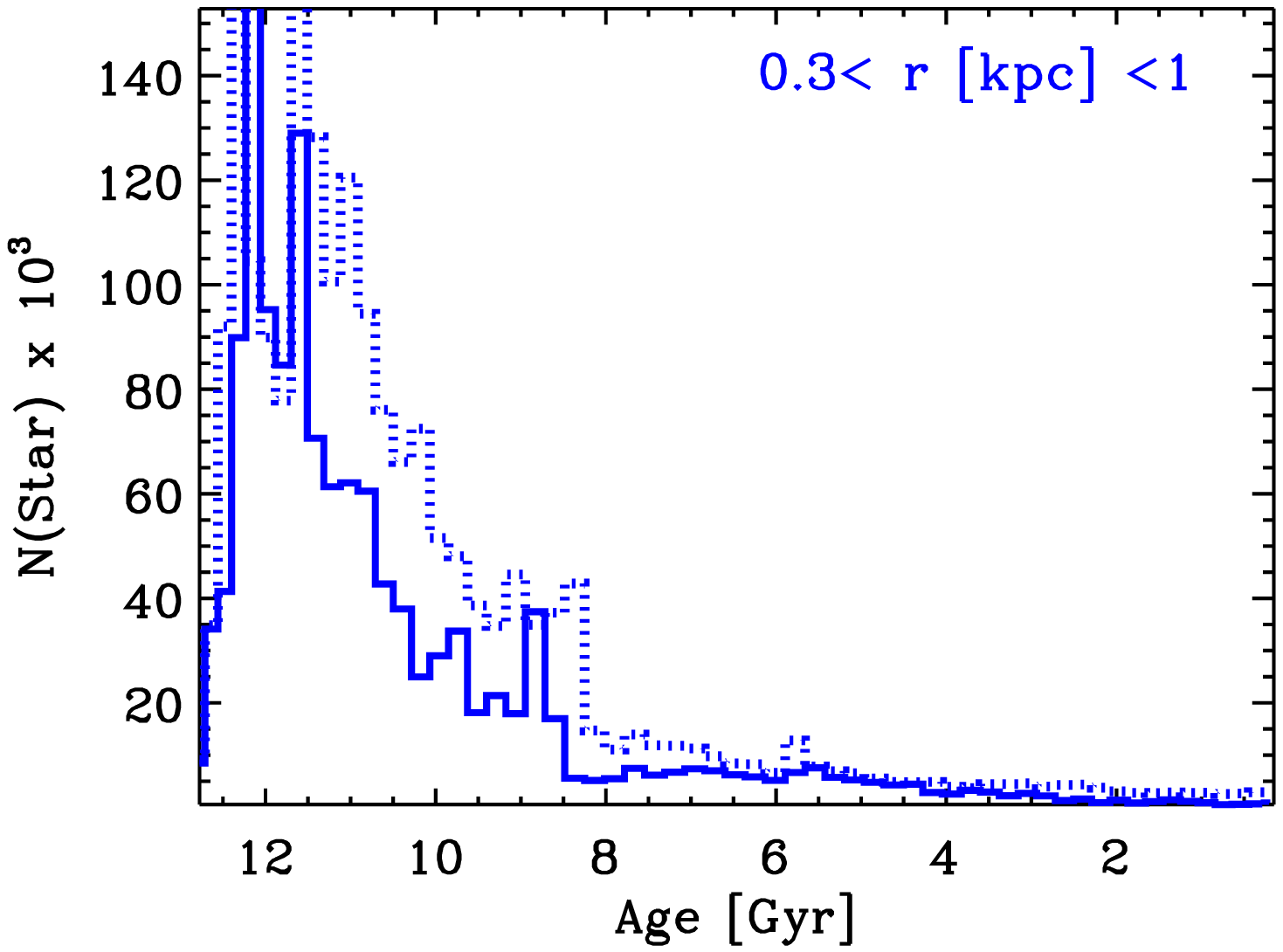}
        \includegraphics[width=0.23\textwidth]{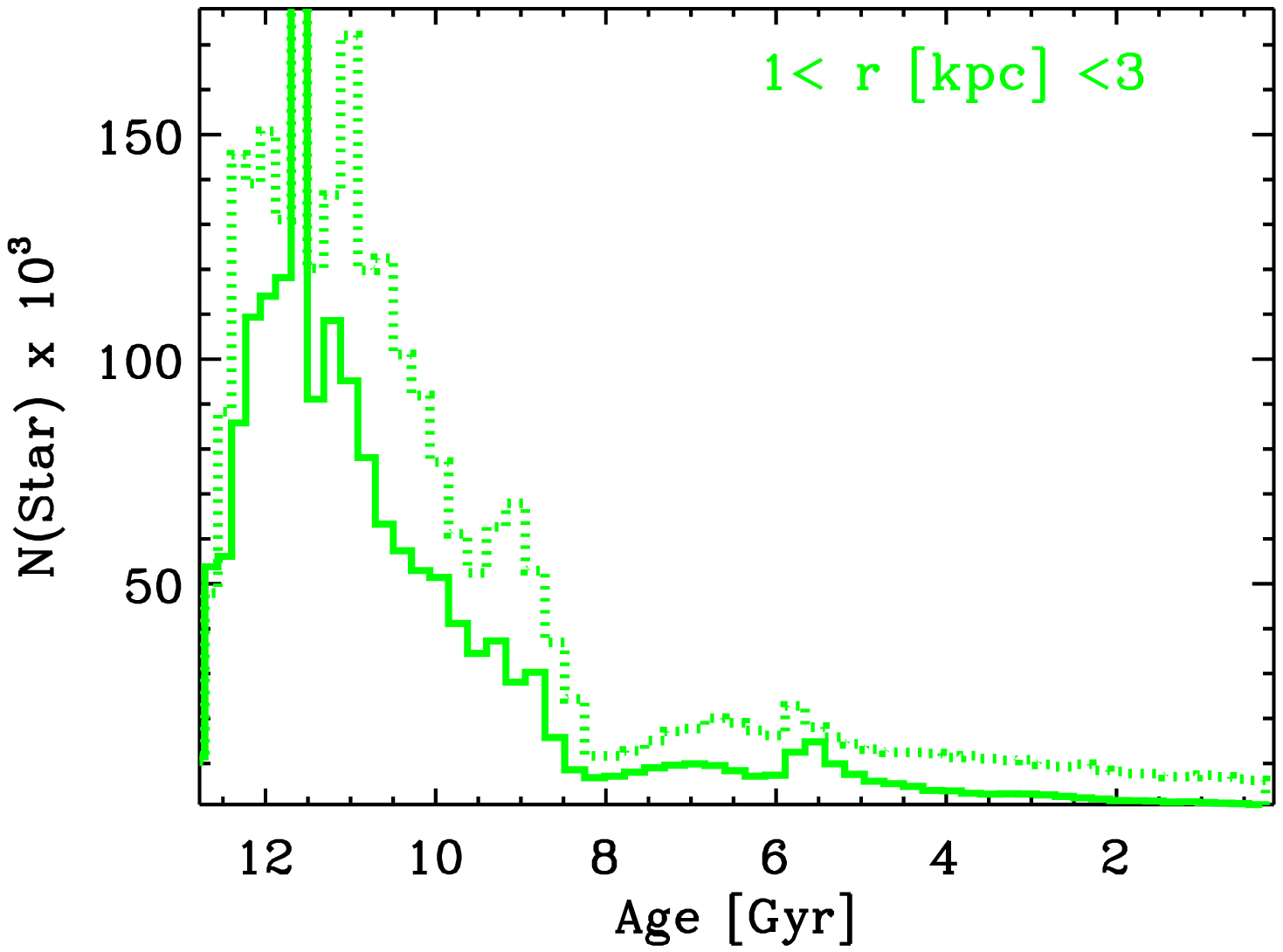}
        \includegraphics[width=0.23\textwidth]{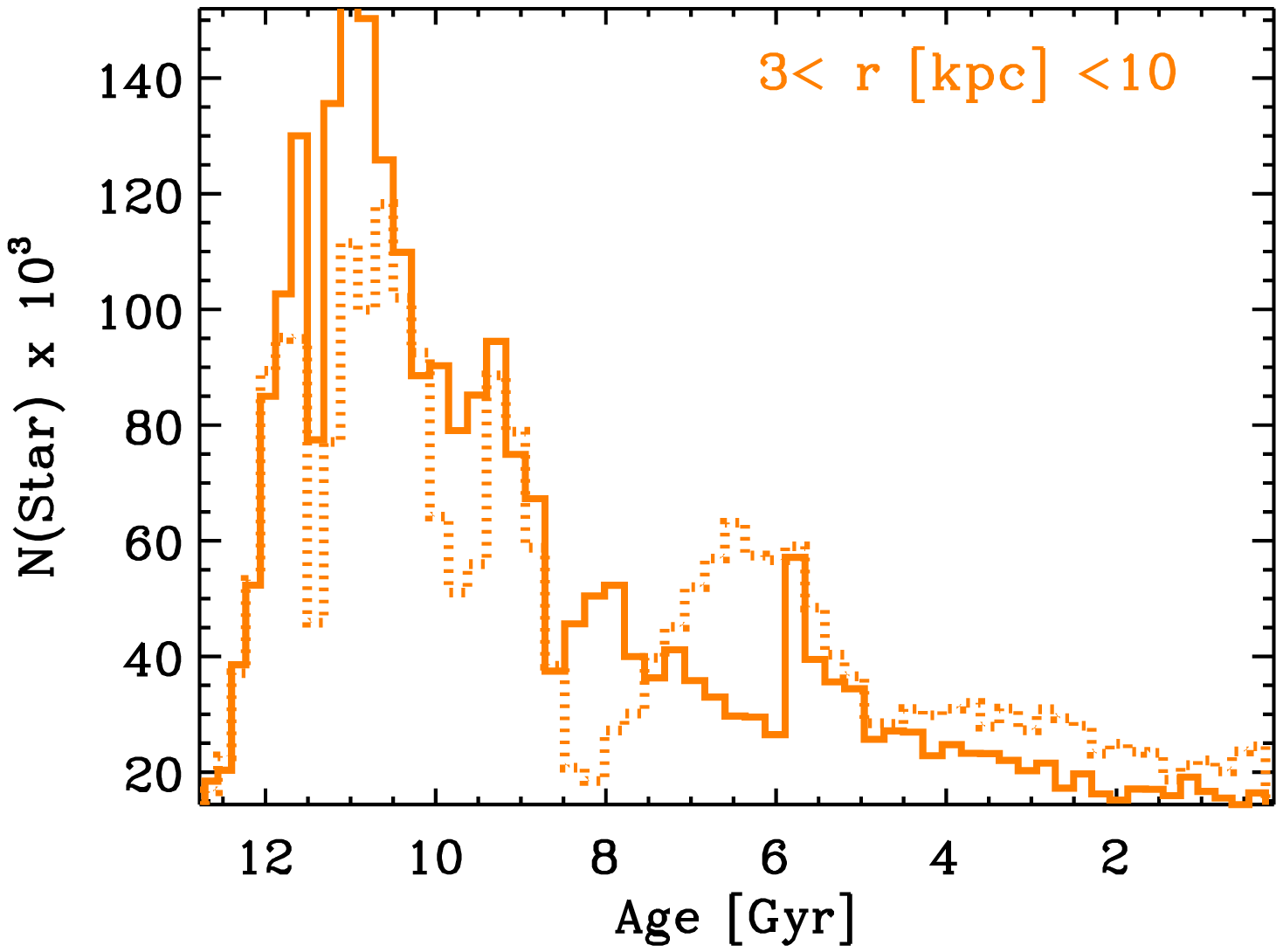}
        \caption{Age distribution of stars located, at $z=0$, in  shells with
radial distance from the galaxy center  as defined in the caption. Solid lines are for \erisbh and
 the dotted ones for Eris. }
        \label{fig:stellar_ages}
\end{center}
\end{figure}

As discussed in details in \citet{guedes13}, the pseudobulge of Eris is the
result of an inside-out growth, mainly evolved from a stellar bar that formed at
high redshift and that has progressively weakened after $z=1$. In \erisbh
the evolution at recent times is different: the weak AGN feedback has inhibited
the growth of the bulge, while its disk has actually grown to a
larger size with respect to Eris (see the right panel of
\ref{fig:Mass_enclosed_ph} for quantitative differences in the stellar content of
the two simulations at various radii). Being the disk larger and the bulge smaller
(and thus with less stabilizing power), the disk of \erisbh is more prone to
instabilities. While the bar of Eris thus progressively weakens after $z=1$, in
\erisbh the disk becomes again unstable at $z \sim 0.3$\footnote{the ratio
$v_{\rm peak}/\sqrt{G
M_{\rm d}/R_{d}}$ (where $v_{\rm peak}$ is the peak of the circular velocity and
$M_{\rm d} and R_{d}$ are the mass and scale-length of the disk) is
traditionally used to estimate the stability of galactic disks \citep{mo98}.
Simulations show that a disk becomes unstable when the ratio becomes smaller
than
$1.1$. We find that, at $z=0.3$, \erisbh has precisely a value of $1.1$, while,
for comparison, Eris has a value of $1.4$. Those numbers are consistent with the disk of \erisbh
being about to become unstable, while Eris does not suffer any instability event.} and the bar gains again
strength with several consequences for the stellar and gas properties (e.g.,
Figures \ref{fig:Gas_map}, \ref{fig:sfr_comp}, \ref{fig:stellar_ages}).

\begin{figure}
\begin{center}
        \includegraphics[width=0.45\textwidth]{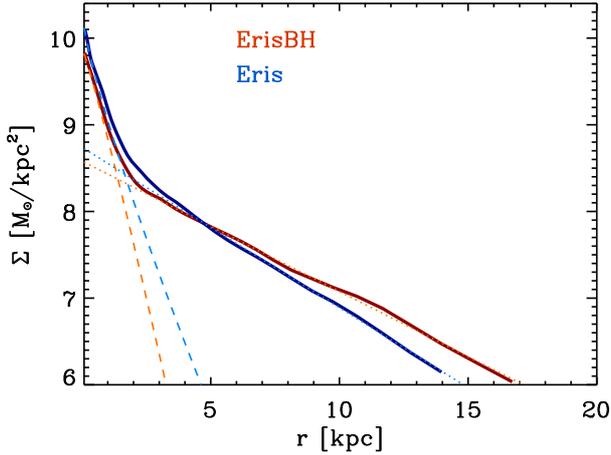}
        \caption{Stellar density profiles of \erisbh  and Eris at $z=0$ (solid
lines). The two-components  S\'ersic fits are shown by the dashed and dotted curves,
for the bulge and disk, respectively. The values of the best-fit parameters are
given in Table \ref{table:info_global}.}
        \label{fig:dens_profile_z0_fits}
\end{center}
\end{figure}

The stellar density profiles of \erisbh and Eris are shown in Figure
\ref{fig:dens_profile_z0_fits}, together with the best-fits for the bulge and
disk components. The profiles have been fitted with a double S\'ersic profile, with
the S\'ersic index for the bulge left unconstrained, and the one for the disk
fixed at $1$ (exponential disk). The best values for the fits are given in Table
\ref{table:info_global}. The bulge of \erisbh  features a smaller S\'ersic index
than Eris ($n_{\rm s} = 0.9$ and $n_{\rm s} = 1.2$, respectively), with a value
lower than unity, typical of boxy bulges.  The scale radius of the disk of
\erisbh is instead larger than the one of Eris ($R_{\rm disk} =2.9$ and $R_{\rm
disk} =2.3$, respectively). We obtained the total bulge and disk masses by integrating
the S\'ersic profiles, and got the values also given in Table
\ref{table:info_global}: the bulge of \erisbh is almost half the one of Eris,
while the disk is slightly larger. As discussed above, the weak AGN feedback is 
responsible for the lower reservoir of gas in the central few hundred parsecs of \erisbh 
compared to Eris and the consequent lower star formation rate and stellar
content in the central region of the galaxy (see Figures \ref{fig:sfr_comp} and  \ref{fig:Mass_enclosed_ph}).
We note that the values we obtained here for  the S\'ersic index of the bulge and the scale
radii for Eris differ from the ones given in \citep{guedes11}. There, the fit
was performed with the GALFIT code run on the surface brightness profile
obtained after processing the simulated galaxy with SUNRISE. Using this methodology, \citet{guedes11} obtained
a photometric $B/D$ ratio of $0.35$ in the B-band, lower of about a factor of
$2$ than the $B/D$ ratio we
obtain here. Assuming that for \erisbh the ratio between the photometric
$B/D$ ratio and the $B/D$ ratio as we calculated here is the same as for Eris
(which is plausible, given the similar stellar ages at
 galaxy scales between the two simulations, see Figure \ref{fig:stellar_ages}),
 we obtain an estimated value for the photometric B-band $B/D$ ratio 
of $0.19$ for \erisbh, which would classify the galaxy as a barred $Sc$ spiral.

We now look at the differences in sfr history between the two simulations. The
evolution of the  sfr of Eris and \erisbh if shown in
Figure \ref{fig:sfr_comp} (empty and filled squares respectively). The symbols
of different colors show the sfr averaged over $10$ snapshots (approximately
$300 \Myr$) for the entire box
(black) and within spheres of different radii from the center. The largest
differences between the two simulations are in the inner
few $\kpc$, where the star formation rate can be up to an order of magnitude
lower in \erisbh than in Eris. This is reflected in the stellar distribution, which is shown in
Figure \ref{fig:Mass_enclosed_ph}: within few $\kpc$ the stellar content in \erisbh
 can be up to $50 \%$ lower than in Eris, while above $10 \kpc$ there are no
striking differences between the two simulations. 

Apart for being generally lower in the central regions, the sfr in \erisbh
 has also a different redshift evolution: while in Eris the sfr
 gradually decreased after $z=1$, in \erisbh we see still some small bursts of
star formation at more recent times. There is
particularly a new increase of star formation at $z <0.3$, after the appearance
of the strong nuclear bar visible at $z=0$. It is likely that the bar is
driving a new inflow of gas to the center (see also the gas distribution in the
central $\kpc$ in the left panel of Figure \ref{fig:Mass_enclosed_ph}) which leads
to a new episode of star formation.

The presence of a young generation of stars in the very center of \erisbh can be
seen also in Figure \ref{fig:stellar_ages}, where we show the age distribution
of the stars within different radial shells from the center.  The very central
stars (within $300 \pc$)
of \erisbh (solid histogram on top panel) are primarily quite old, except
for a population of young stars ($age < 2 \Gyr$) originating from the new
episode of
star formation at $z<0.3$ which takes place during the formation of the bar. In
Eris (dotted histogram) the age
distribution in the inner region is different: there is a large population of
stars born during the last merger at $ z \sim 1$ ($age \sim 7 \Gyr$), and a
slowly decreasing distribution at younger ages  (see the corresponding increase
in the sfr in
the inner region around $z  \sim 1$  in Figure \ref{fig:sfr_comp}, followed by a
gradual decrease of the sfr). 

The role of bars in bringing a population of young stars within the  bulges of
disk galaxies has been studied by \citep{coelho11}, where the authors speak
about a ``rejuvenation'' of bulges due to the presence of bars.

\begin{figure}
\begin{center}
        \includegraphics[width=0.2\textwidth]{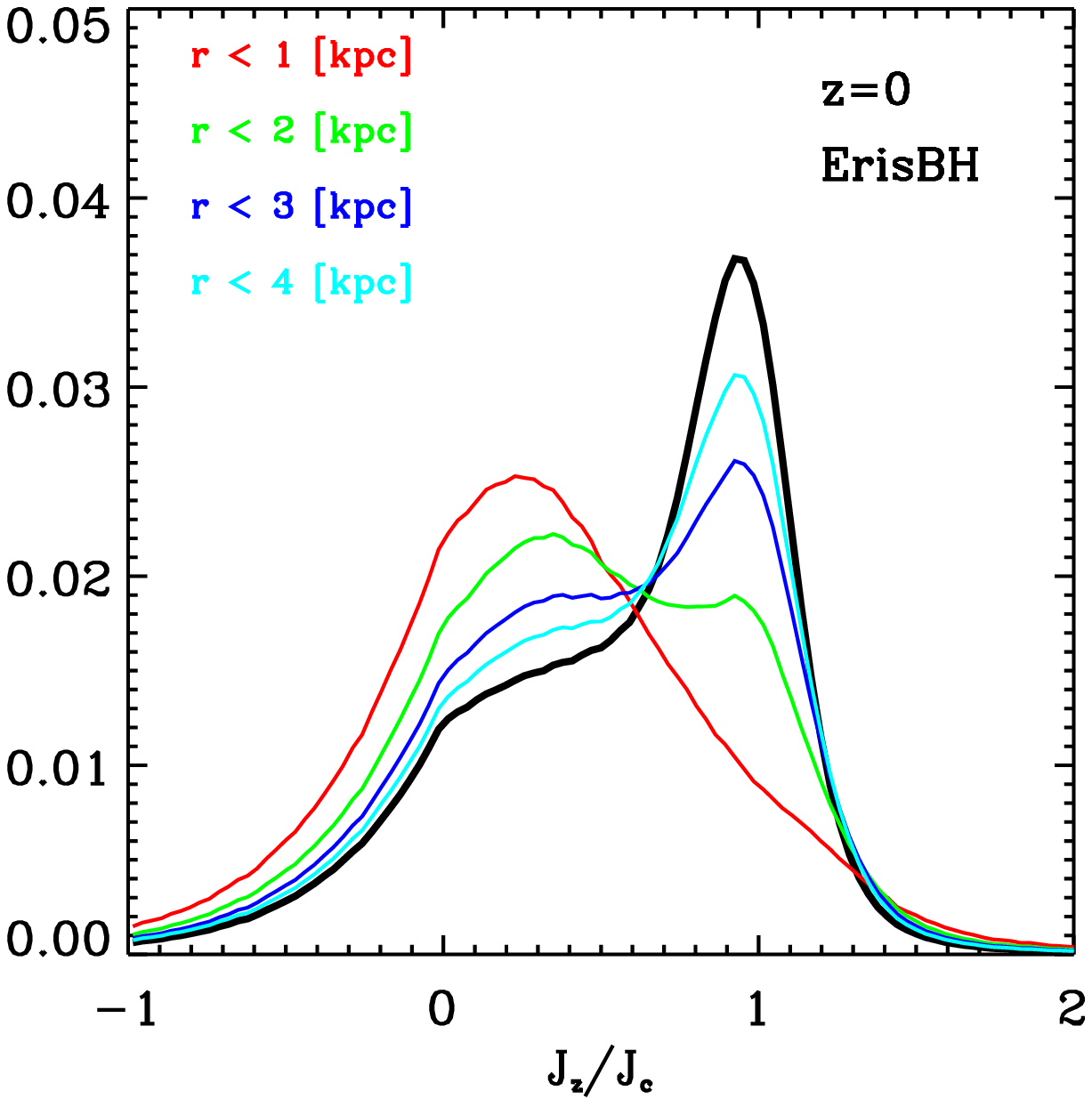}
        \includegraphics[width=0.2\textwidth]{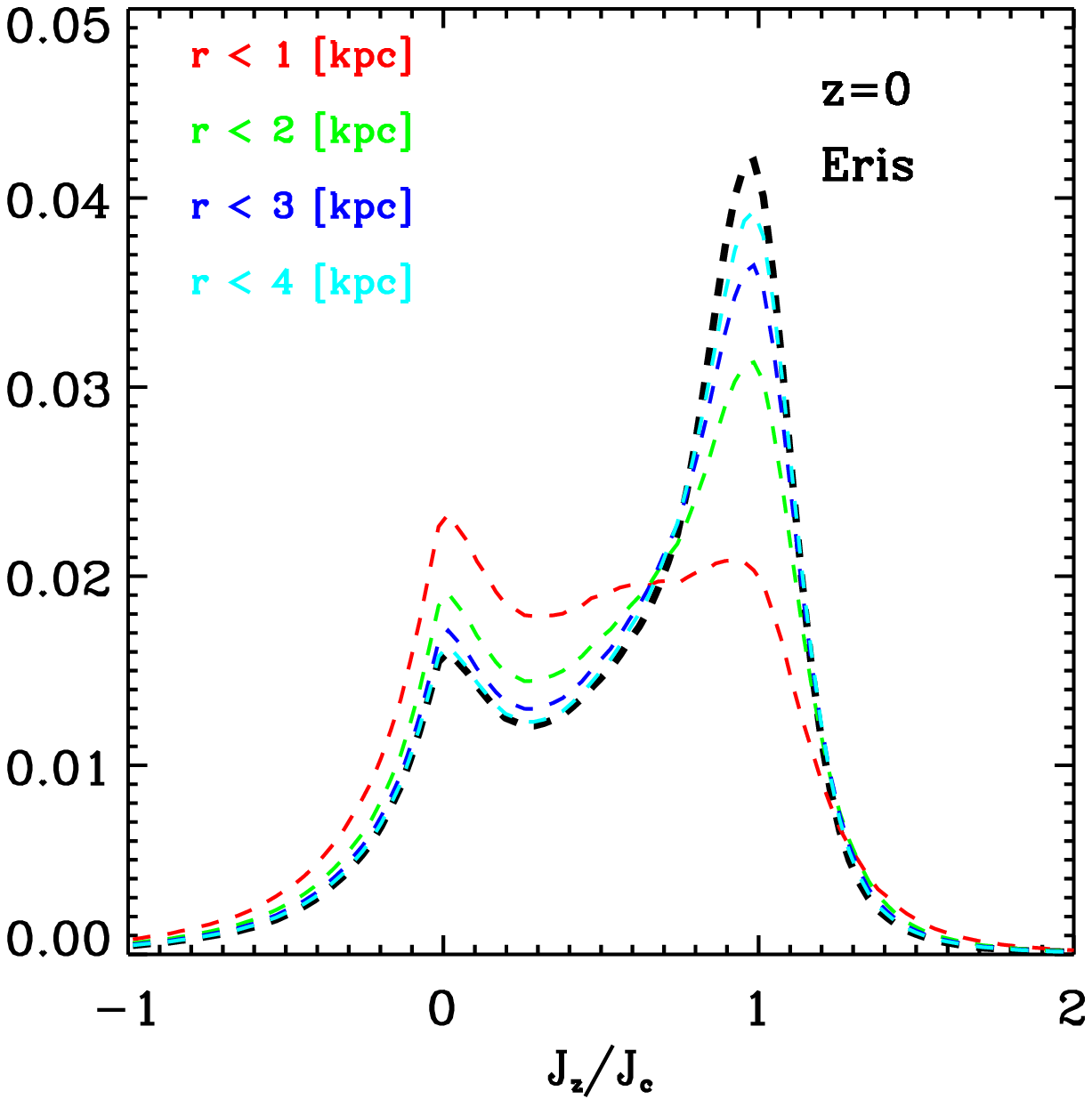}
        \caption{Distribution of ``orbital circularity parameter'' $\epsj$ for
 \erisbh (left panel) and Eris (right panel) at $z=0.3$ . The distribution including all stars within  $20 \kpc$ from the center
is given by the black lines. The colored lines show the distribution for stars
 within spheres with radial distance from the center as indicated in the
caption.}
        \label{fig:epsj}
\end{center}
\end{figure}

On top of causing a new inflow of gas which leads to the new burst of star
formation, the bar has also a dynamical effect in the distribution of angular
momentum in the center of the galaxy.
Figure \ref{fig:epsj} shows the distribution of the ``orbital
circularity parameter'' $\epsj$\footnote{The orbital circularity parameter is
used to kinematically separate stars belonging to the disk and stars belonging
to the bulge. It is defined as 
 $\epsj = j_{\rm z}/j_{\rm circ}$, where $j_{\rm z}$ is the angular momentum of
each star in the z-direction (i.e. the direction of $J$), and $j_{\rm circ}$ is
the angular momentum expected for a circular orbit at the same radius.} for \erisbh (left) and Eris (right)
at  $z=0$ for all stars within $20 \kpc$ from the center (black curves) and for
stars within different radii from the center (colored lines). 
 Starting from the
very central $\kpc$ (red curves), there is a striking difference between \erisbh
and Eris: in Eris (right panel) there is a narrow peak
of stars around $\epsj =0$, and a significant fraction of higher angular
momentum particles that reach the central region. This is not happening in
\erisbh, where no high-angular momentum stars
 are present in the center of the galaxy, consistent with the presence of a
strong bar. The detailed effects of the bar on the dynamical properties of the
galaxy will be studied in a
 separate paper.

\begin{figure}
\begin{center}
        \includegraphics[width=0.45\textwidth]{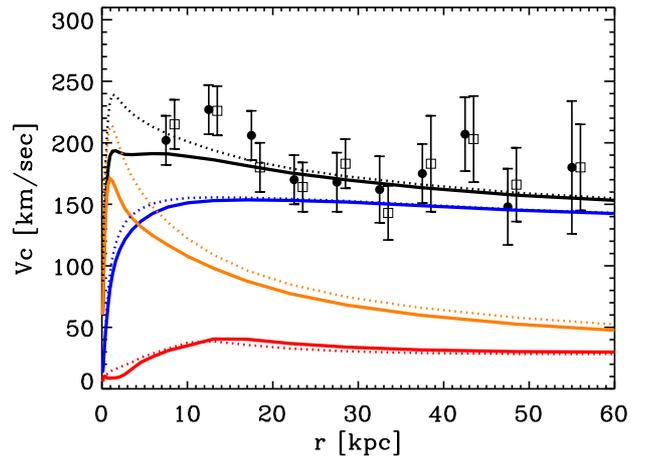}
        \caption{Rotation curves  for
\erisbh (solid lines) and Eris
(dotted lines) in the inner $60 \kpc$.   Total vc (black curves), for dark
matter only (blue curves), for stars
only (orange curves), for gas only (red curves). The data points are from
\citet{xue08}. }
        \label{fig:vc}
\end{center}
\end{figure}

Finally, the lower stellar content in the center of \erisbh results to a lower rotation
curve, with respect to Eris, in the inner $10 \kpc$ (Figure \ref{fig:vc}). A
value around $190-200 \kms$ for the peak of the circular velocity in the inner
region of the galaxy is
 consistent with the values 
estimated for the Milky Way by \citep{portail15}, who constructed dynamical
models of the bulge of the Galaxy using the 3D density measurements of Red Clump Giants and kinematic data from the
BRAVA survey.
At  large distances, on the other side,  the rotation curves
of the two simulations are almost identical, and
 consistent with the data of \citet{xue08}, who estimated the Milky Way's rotation curve  using the kinematics of blue horizontal-branch stars.

To conclude, despite being quantitatively modest, the feedback from the central
black hole of \erisbh leads to  small changes in the gaseous and stellar
distribution in the inner region of the galaxy, when compared to the Eris
simulation  (where AGN feedback was neglected). Those small variations are then
the origin of  important
secondary effects, particularly in the stability of the galaxy. 
\erisbh, in fact, experiences a  
strong instability at $z\sim 0.3$, that leads to the formation of a new large
bar.  This event causes the  depletion of gas in the inner few
kiloparsecs, and further growth of the bulge is inhibited. By $z=0$, the central
galaxy of \erisbh has a larger disk and a smaller bulge, with a box/peanut
morphology, with respect to  the Eris galaxy.

\begin{table*}
 \begin{center}
  \begin{tabular}{c||cc|cc|cc|cc|cc}
	\hline
	  &  $M_{\rm star}$  & $M_{\rm gas}$  &
$f_{b}$  & $V_{\rm max}$ & $\MDisk$  & $\MBulge$ & B/T & $R_{\rm disk}$ &
$R_{\rm bulge}$ & $n_{\rm s} $ \scriptsize{(Bulge)}\\
        \hline
        \hline
        \erisbh &  $3.2 \times 10^{10}$  &
$5.9 \times 10^{10}$  & $0.118$ & $193.2$  & $2. \times 10^{10}$ & $8.4 \times
10^{9}$& $0.29$  & $2.9$ & $0.4$  &  $0.9$ \\
        \hline
        Eris &  $3.9 \times 10^{10}$  &  $5.6
\times 10^{10} $  & $0.121$ & $239.2 $ & $1.9 \times 10^{10}$ & $1.5 \times
10^{10}$ & $0.44$ &$ 2.3$ & $0.3$ &  $1.2$ \\
        \hline
        \hline
  \end{tabular}
  \label{table:info_global}
 \caption{Summary of global properties of \erisbh and Eris at $z=0$. The
stellar and gas masses in columns $1$ and $2$ refer to the total
stellar and gas masses within the
halo, also used to calculate the baryon fraction given in column $3$. The
disk and bulge masses have been obtained fitting a double S\'ersic profile to the
 face-on surface density profile, with scale lengths given in columns $8$ and
$9$. The last column gives the S\'ersic index for the bulge, while the S\'ersic
index for the disk has been fixed to unity.  Masses are
given in solar masses, distances in $\kpc$ and velocities in $\kms$. }
 \end{center}
\end{table*}

\section{Summary and Conclusions} \label{sec:conclusions}
In this paper we presented \erisbh, a new zoom-in cosmological simulation where
we followed the evolution of a Milky Way-size dark matter halo with its baryon
content from $z \sim 90$ to the present time. \erisbh  is
a twin simulation of Eris \citep{guedes11}, with which it shares the same
initial conditions,  simulated physical processes and the resolution of $120 \pc$, but,
differently from Eris, it includes also
prescriptions for the formation and evolution of massive black holes. 

Four black holes are inserted during the simulation run in the 
 most massive galaxies that also contain a large gaseous high-density region. The largest
black hole, located in the
largest galaxy of the simulated volume, is seeded at $z \sim 8.5$ with an
initial mass of about $9 \times 10^5 \Msun$.

In this first paper on \erisbh we focused on
studying the properties of this central black hole, its growth, and its relation
and influence on the host galaxy. Our main conclusions are:
\begin{itemize}
\item{During its evolution, very little gas reaches the center of the main galaxy,
thus little fuel is available for feeding the main black hole, whose growth is
generally highly sub-Eddington. Ending up at
$z=0$ with a mass of $2.6$ million Solar
masses, only $\sim 3$ times its initial value, the black hole grows
primarily (about a factor of two) through mergers with the infalling black
holes originally hosted by satellite galaxies;  }
\item{ Given the little gas supply, the value of the initial seed mass is
rather important. Generalizing our results, we  expect Milky Way-size
galaxies to be hosting an intermediate-mass black
hole already at $z \sim 8$. This could be possible either via a direct collapse
in the protogalaxy, or  via the uninterrupted growth close to
the Eddington rate of a Pop
III star remnant. }
\item{The final black hole sits on the $\MBH-\sigma$ and $\MBH-\MBulge$ relations
in a location close to the one of SgrA*. Consistent with
 observational results for pseudobulges, \erisbh is close to the extrapolation
 at lower masses of the $\MBH-\sigma$ relation established by more massive
galaxies, while it is almost an order of magnitude below the $\MBH-\MBulge$ relation; }
\item{Given the little growth by accretion of the central black hole, the AGN feedback on the galaxy is quite 
modest. Only the very central region of the galaxy (within $\sim 1-2 \kpc$) seems to
be affected by the presence of the black hole, with the consequence
that the gas content and the
star formation rates in the central region of the galaxy are
lower than the ones found in Eris;}
\item{Because of its smaller bulge and the lower concentration of gas in the
center, the disk of \erisbh is more unstable. The last instability event, at $z
\sim 0.3$, causes the formation of a bar of about $3 \kpc$ in radius and a burst of star formation in
the very central few hundred parsecs;}
\item{At $z=0$,  the disc of \erisbh is slightly larger than the one of Eris,
while its bulge is almost a factor of two smaller. Consequently, the B/D ratio
of \erisbh is about half of the one of Eris and its rotation curve
flatter. Moreover, the
 bulge of
\erisbh is characterized by the box/peanut morphology typical of barred galaxies, and, at its very
center, it hosts a population of very young stars.}
\end{itemize}

In summary, as expected, progenitors of late-type galaxies seem to not be favourable sites for
efficient black hole growth,  probably due to their rather quiet
merger history. While we do no witness any dramatic episode of
AGN feedback, we find, however, that the weak, but circumscribed, energy input
from the black hole helps the
global stellar feedback to inhibit the growth of the bulge, which has important
consequences on the global stability of the disk. Modest-size black holes,
though, might play an important role in the dynamical evolution of late-type
spirals.

\section*{Acknowledgments}
We thank Massimo Dotti, Dimitri Gadotti and Ortwin Gerhard for useful discussions.
This work was  possible thanks to the computing time allocations at the Swiss National
Supercomputing Center and at the
Ohio
Supercomputer Center.  PM
acknowledges support by the NSF through grant AST-1229745 and by NASA
through grant NNX12AF87G. FG acknowledges support from NSF grant AST-0607819 and
NASA ATP NNX08AG84G. 

\bibliographystyle{mn2e}
\bibliography{ErisBH}

\label{lastpage}

\end{document}